\documentclass[a4paper,11pt]{article}

\usepackage{booktabs,longtable,xcolor,colortbl}

\usepackage{jheppub,amsthm} 

\usepackage[utf8]{inputenc}

\usepackage{booktabs,multirow, quotes}
\usepackage{chngcntr}
\usepackage[english]{babel}
\usepackage{amsmath,amssymb,graphicx,xcolor,alltt,tikz,subcaption}
\usepackage{framed}

\theoremstyle{remark}

\theoremstyle{theorem}

\makeatletter
\def\@fpheader{\ }
\makeatother

\counterwithout{figure}{section}
\title{Field theory guides Monte Carlo in the one-dimensional long-range Ising model }

\author{Ant\'onio Antunes$^\phi$, Jos\'e Matos$^\sigma$}

\affiliation{$^\phi$Laboratoire de Physique de l'\'Ecole Normale Sup\'erieure, Universit\'e  PSL, CNRS, Sorbonne Universit\'e, Universit\'e  Paris Cit\'e, 24 rue Lhomond, F-75005 Paris, France\\
$^\sigma$Laboratory of Physics for Materials and Emerging Technologies (LaPMET), Centro
de F\'isica das Universidades do Minho e Porto (CF-UM-UP),
Departamento de F\'isica e Astronomia, Faculdade de Ci\^encias, Universidade do
Porto,
Rua do Campo Alegre s/n, 4169-007 Porto, Portugal.\\
Fakult\"at f\"ur Physik, Universit\"at Bielefeld, Bielefeld, D-33615, Germany}

\emailAdd{antonio.antunes@ens.fr}
\emailAdd{jose.bouradematos@gmail.com}

\abstract{The statistical long-range Ising (LRI) model in one dimension provides a family of some of the simplest non-exactly solvable critical points. These critical points are described by long-range conformal field theories (CFTs) that are well understood analytically with perturbative expansions and some exact results. Guided by this, we revisit the Monte Carlo (MC) approach to the corresponding lattice model which has many favorable features compared to other systems studied with MC methods. In particular, with very modest computational cost, we are able to estimate scaling dimensions of leading and subleading operators as well as the associated three-point functions, and even certain four-point functions. These results help highlight the fact that 1d LRI CFTs are a phenomenal playground to more generally test non-perturbative methods and ideas for strongly coupled CFTs.}

\begin{document} 
	
	\maketitle
	
	\flushbottom

\section{Introduction}
\label{sec:intro}

The understanding of second order phase transitions has been deeply influenced by the modern conformal bootstrap \cite{Rattazzi:2008pe,ElShowk2012,Kos2016}. Both its numerical incarnation, which predicts critical exponents with high precision and mathematical rigor \cite{Rattazzi:2008pe,Kos2016,Reehorst:2026vqe}, and its analytic version which describes structural properties of the spectrum of local operators and their correlation functions \cite{Komargodski:2012ek,Fitzpatrick:2012yx,SimmonsDuffin2017,CaronHuot2017}, have greatly benefited from the existence of an interacting and non-exactly solvable universality class: that of the 3d Ising conformal field theory (CFT), which has become a general testing ground for non-perturbative methods both on the lattice and in the continuum.

In many contexts, the 3d Ising CFT is treated as the simplest non-exactly solvable conformal field theory. This is in contrast to most well understood 2d CFTs which are exactly solvable\footnote{See however \cite{DotsenkoJacobsenLewisPicco1999,AntunesBehan2023,AntunesBehan2025,AntunesRong2026,AntunesSuchel2025} for proposals of non solvable 2d universality classes} and local one-dimensional CFTs which are trivial \cite{Ruelle1968}. Crucially, relaxing locality decreases the lower critical dimension without sacrificing conformal invariance of the RG fixed points \cite{PhysRevLett.29.917,PaulosRychkovVanReesZan2016}. In particular, the one-dimensional long-range Ising model, provides a family of interacting 1d CFTs that cannot be exactly solved \cite{BenedettiLauriaMazacVanVliet2025,BenedettiLauriaMazacVanVliet2026}. As we will review in section \ref{sec:review}, the assymptotic density of operators is comparable to that of a local 2d CFT, making the system intrinsically simpler than the 3d Ising CFT, not to mention that the absence of spacetime spin in 1d dramatically simplifies the kinematics. We therefore advocate that\footnote{We thank Miguel Paulos who strongly influenced our view of the subject.}

\begin{center}
    \textit{The interacting fixed points of the one-dimensional long-rang Ising model are the simplest non-exactly solvable CFTs and should be used as a benchmark for nonperturbative methods in QFT and lattice models.}
\end{center}

 The 1d LRI CFTs have already received a moderate amount of attention in the literature, including perturbative expansions around their mean-field and short-range ends \cite{PhysRevLett.29.917,BenedettiGurauHarribeySuzuki2020,BenedettiGurauHarribey2025,BehanRastelliRychkovZan2017,BenedettiLauriaMazacVanVliet2026}, bootstrap studies \cite{Behan:2018hfx, BenedettiLauriaMazacVanVliet2025,BehanLauriaNocchiVanVliet2024,GhoshPaulosSuchelZheng2026} and Monte-Carlo simulations \cite{LuijtenBlote1997,AngeliniParisiRicciTersenghi2014,FUKUI20092629,PhysRevE.63.037101,4e6680214fdb4bf18165a5331ee0dc79}.\footnote{Other methods, such as Hamiltonian truncation also seem ripe for application in this context \cite{PaulosCardinaleWIP}.} However, we found that the available MC studies in the literature are not sufficiently targeted on the interesting observables for the CFT community and therefore aim to address this gap.

 In particular, the continuum description of the model in terms of a perturbed generalized free field (GFF) \cite{PaulosRychkovVanReesZan2016,BenedettiLauriaMazacVanVliet2025} provides many useful insights which help improve the MC simulations including the existence of a protected scaling dimension for the order parameter as well as the behavior under RG away from the interacting fixed point \cite{PaulosRychkovVanReesZan2016,BehanRastelliRychkovZan2017}. Furthermore, there are many favorable features for the study of 1d LRI CFTs using MC methods. The low scaling dimension of the order parameter ensures many observables vary rather slowly and there exist highly efficient algorithms whose computational complexity is the same as for local one-dimensional models in spite of the non-local interactions. 
 
 Making use of these insights, we explore our proposal of the 1d LRI as a testbed for non-perturbative approaches, studying a plethora of observables with very modest computational resources (sometimes only hundreds of CPU hours). We make a precise determination of the critical temperature for several values of the long-range decay exponent; determine the scaling dimension of the leading energy operator in several independent ways; access subleading scaling dimensions; estimate OPE coefficients and even conformal four-point functions. 

\medskip 

The rest of this paper is organized as follows: In Section \ref{sec:review} we review the construction of the LRI model both in the continuum and on the lattice, and recall the relevant MC algorithms. In Section \ref{sec:MC}, we study the determination of the critical temperature using both standard and field-theory inspired methods and consider some additional off-critical properties of the model. We obtain our first estimate of the energy operator scaling dimension in Section \ref{ssec:thermal-response} by studying thermal fluctuations of the spin-spin correlator. We study more general correlation functions involving the energy operator and its subleading counterparts as well as higher-point functions with spin operators in \ref{sec:results}.
The main results of the paper, concerning the scaling dimension of the leading and subleading energy operator as well as some three-point function coefficients are summarized in Section \ref{sec:comparison}. We end with a brief discussion of future directions in Section \ref{sec:discuss}. Additional results on the determination of the critical temperature are presented in Appendix \ref{app:all critical couplings finite volume dependence}.

\section{Brief review of LRI}
\label{sec:review}
In this section we will review some key concepts and results on the 1d LRI model in the continuum and on the lattice. In section \ref{ssec:QFT} we recall the QFT construction of the model in terms of a generalized free field deformed by quartic interactions, emphasize some relations to AdS physics and quote key perturbative results \cite{PaulosRychkovVanReesZan2016,BenedettiGurauHarribeySuzuki2020,BenedettiGurauHarribey2025}. In Section \ref{ssec:lattice} we specify our conventions for the lattice model with periodic boundary conditions and review the powerful Fukui-Todo algorithm \cite{FUKUI20092629}.
\subsection{Field theory construction}
\label{ssec:QFT}
The continuum limit of the lattice LRI Hamiltonian is given by the $\mathbb{Z}_2$-symmetric action
\begin{equation}
    S_{\rm{LRI}}[\phi]= S_{\rm{GFF}(\Delta_\phi)}[\phi]+\int dx \left(\frac{\lambda_2}{2}\phi(x)^2 + \frac{\lambda_4}{4}\phi(x)^4\right)\,,
\end{equation}
with the GFF $\phi$ being described by the bi-local Gaussian action
\begin{equation} 
    S_{\rm{GFF}(\Delta_\phi)}[\phi]= \int dx_1 dx_2 \frac{\phi(x_1)\phi(x_2)}{|x_1-x_2|^{1+s}}\,,
\end{equation}
where there is an exact relation between the parameter $s$ and the scaling dimension of the GFF $\Delta_\phi$
\begin{equation}
    \Delta_\phi=\frac{1-s}{2}\,,
\end{equation}
ensured by the locality of the interactions which cannot renormalize the non-local kinetic term \cite{PaulosRychkovVanReesZan2016}. This relation will play a key input in the determination of the critical temperature in Section \ref{ssec:critfromprotect}. 

The non-local nature of the action does not make manifest the number of degrees of freedom of the CFT, i.e. the density of states in radial quantization. This is better understood using the description of GFF in terms of a free field $\Phi$ in two-dimensional Anti de Sitter/hyperbolic space. An elementary exercise in holographic renormalization shows that \cite{KlebanovWitten1999,PaulosRychkovVanReesZan2016}
\begin{equation}
     S_{\rm{GFF}(\Delta_\phi)}[\phi] \Leftrightarrow S_{\mathrm{AdS}_2(m)}[\Phi]\,,
\end{equation}
with 
\begin{equation}
    S_{\rm{AdS}_2(m)}[\Phi]= \int_{\rm{AdS}_2} d^2x \sqrt{g} \left(\frac{1}{2} (\partial_\mu\Phi)^2+ \frac{1}{2} m^2  \Phi^2\right)\,,
\end{equation}
with the standard AdS/CFT relation \cite{KlebanovWitten1999}
\begin{equation}
\label{adscftdict}
    m^2=\Delta_\phi(\Delta_\phi-1)\,,
\end{equation}
in units where the AdS radius $L$ is set to one. The state-operator correspondence for QFT in AdS$_2$, and in particular for the two-dimensional free field, then ensures the 1d GFF density of states $\rho(\Delta)\sim \exp(\sqrt\Delta)$, comparable to a local 2d CFT.

Returning to the AdS/CFT dictionary \eqref{adscftdict}, there are two choices for $\Delta_\phi$ for a given $m^2$ \cite{KlebanovWitten1999} corresponding to Dirichlet $\Delta_{+}=(1+\sqrt{1+4m^2})/2$ or Neumann $\Delta_-=1-\Delta_+$ boundary conditions. The relevant choice for LRI, corresponds to the Neumann boundary conditions where there are relevant boundary operators. For example, in the range $1/4<\Delta_\phi\leq1/2\,,$ there is a single $\mathbb{Z}_2$ even relevant singlet, the operator $\phi^2$. The associated RG flow, which has no dimensionless parameters, is the famous Neumann to Dirichlet or double-trace flow \cite{Witten2001,GubserKlebanov2003}
\begin{equation}
    S_{\rm{GFF}(\Delta_\phi)} +\lambda_2 \int dx \,\phi^2(x) \rightarrow  S_{\rm{GFF}(1-\Delta_\phi)}\,,
\end{equation}
meaning that adding a mass does not gap the system, it instead changes the power law in the infrared. This will help us understand the off-critical physics of the lattice model to be discussed in Section \ref{ssec:offcrit}.  Continuing towards the range $\Delta_\phi\leq1/4$ introduces a second relevant operator $\phi^4$, and one therefore needs to tune the dimensionless combination of $\lambda_4$ and $\lambda_2$ for each $s$ to find the interacting LRI fixed point.\footnote{In fact, as one continues to lower $\Delta_\phi$ more and more $\phi^{2n}$ operators become relevant, suggesting the existence of an infinite family of long-range multi critical fixed point. See \cite{BehanBenedettiEustachonLauria2026} for a 2d analogue of these models. } 

For $1/2<s<1$, the non-local equation of motion at the interacting fixed point $\square^{s/2} \phi=\lambda_4 \phi^3$ ensures the non-renormalization of the $\phi^3$ operator\footnote{In fact, the LRI CFTs have infinite families of protected operators due to the higher-spin symmetry of the bulk theory. See \cite{Behan:2018hfx,BenedettiLauriaMazacVanVliet2025,BenedettiLauriaMazacVanVliet2026,Antunes:2025iaw} for relevant discussions.} which satisfies \cite{PaulosRychkovVanReesZan2016,BehanLauriaNocchiVanVliet2024}
\begin{equation}
    \Delta_{\phi^3}=1-\Delta_\phi\,.
\end{equation}
Therefore the two lightest $\mathbb{Z}_2$ odd operators have exactly known scaling dimensions. In fact, their dimensions form what is sometimes known as a shadow-pair as they add-up to the spacetime-dimension $\Delta_{\phi^3}+\Delta_\phi\,=1$. Associated to this relation is a conformally invariant integral transform, known as the shadow transform $\mathcal{S}$ which acts in general spacetime dimension $d$ as \cite{Simmons-Duffin:2012juh}
\begin{equation}
    \mathcal{S}[\mathcal{O}(x)]=\int d^dy\,\frac{\mathcal{O}(y)}{|x-y|^{2(d-\Delta)}}
\end{equation}
and transforms conformally covariant correlation functions of a local primary operator of dimension $\Delta$ to those of a fictitious conformal primary of dimension $d-\Delta$. This scaling dimension is typically unphysical but in the LRI CFTs both members of the shadow paper are part of the spectrum.

\medskip 

Moving on to the $\mathbb{Z}_2$ even sector, the leading non-trivial operator $\phi^2$ is unprotected, and its scaling dimension must be computed in perturbation theory. Around the GFF end, with $\epsilon=2 (s-1/2)$ \cite{BenedettiGurauHarribeySuzuki2020,BenedettiGurauHarribey2025} 
\begin{equation}
    \Delta_{\phi^2}
=\frac{1}{2}- \frac{\epsilon}{6} + \frac{\epsilon^2}{9}(\pi+4\log(2))-3.71975\epsilon^3 + O(\epsilon^4
    )\,.
\end{equation}
The subleading energy operator $\phi^4$ has also been computed to the same order \cite{BenedettiGurauHarribeySuzuki2020,BenedettiGurauHarribey2025}
\begin{equation}
    \Delta_{\phi^4}=1+\epsilon -\frac{2\epsilon^2}{3}(\pi +4 \log(2))+ 48.2275\epsilon^3 + O(\epsilon^4
    )\,,
\end{equation}

In general dimension $d$ the LRI model has a second perturbative region, the so-called short-range crossover region \cite{BehanRastelliRychkovZan2017} where the small coupling parameter is $\delta=\Delta_\phi-\Delta_\phi^{\rm{SR}}$ and $\Delta_\phi^{\rm{SR}}$ is the scaling dimension of the order parameter at the short range fixed point. For example in $d=2$ the close to GFF region is at $\Delta_\phi=1/2$ (since the $\phi^4$ operator becomes classically marginal in $d=2$) while the short range region is near $\Delta_\phi=1/8$: the famously exact 2d Ising CFT dimension. The description \cite{BehanRastelliRychkovZan2017} involves a dual/shadow GFF $\chi$ of dimension $d-\Delta_\phi$ and the formal action is given by 
\begin{equation}
    S_{\rm{LRI}}= S_{\rm{CFT}}^{\rm{SR}} + S_{\rm{GFF}(\Delta_\chi)}[\chi] + \lambda \int d^dx \,\sigma(x) \chi(x)\,,
\end{equation}
where $S_{\rm{CFT}}^{\rm{SR}}$ denotes the formal action of the short-ranged fixed point, and the observables are defined via conformal perturbation theory. In one dimension, where there is no short-ranged fixed point, the description is more subtle. The GFF $\chi$ has a negative scaling dimension and couples to a qubit via two-classically marginal interactions, leading to BKT-like physics \cite{BenedettiLauriaMazacVanVliet2025,BenedettiLauriaMazacVanVliet2026}. In this description, the dimensions of the two lightest $\mathbb{Z}_2$ even operators have been obtained to subleading order \cite{BenedettiLauriaMazacVanVliet2025,BenedettiLauriaMazacVanVliet2026} (once again $\epsilon=2(s-1/2)$ but we now expand it near one). For $\phi^2$ 
\begin{equation}
    \Delta_{\phi^2} =1- \sqrt{1-\epsilon} +\frac{1-\epsilon}{8} + O(\epsilon^{3/2})\,,
\end{equation}
and the subleading $\mathbb{Z}_2$ even operator $\phi^4$ satisfies
\begin{equation}
\label{phi4SR}
    \Delta_{\phi^4} =1 + \sqrt{1-\epsilon} +\frac{1-\epsilon}{8} + O(\epsilon^{3/2})\,.
\end{equation}
Importantly, the expansion is non-analytic near $\epsilon=1$ exhibiting square-root behavior.

\medskip

To obtain robust predictions for these scaling dimensions using the perturbative data, it is convenient to compute a two-sided Padé approximant in the variable $\sqrt{1-\epsilon}$ using all the available orders of both expansions. We computed several approximants with different degrees $[p,q]$ which did not exhibit poles in the region of interest $0\leq\epsilon\leq1$ and that agree among themselves to higher accuracy than needed for the comparison with the MC simulations. We will use the following $[3,3]$ Padés as a proxy for the predictions from the perturbative data
\begin{equation}
    \Delta_{\phi^2}^{\rm{pert}}= \frac{2.52912 \epsilon\sqrt{1- \epsilon } 
   -6.79768 \epsilon -9.83833 \sqrt{1-\epsilon
   }+9.92847}{\epsilon \sqrt{1- \epsilon } 
   -2.22791 \epsilon -5.17842 \sqrt{1- \epsilon
   }+5.3587}\,,
\end{equation}
\begin{equation}
    \Delta_{\phi^4}^{\rm{pert}}= \frac{-0.164819 \epsilon  \sqrt{1- \epsilon }-1.15451 \sqrt{1- \epsilon }+0.0459321 \epsilon +1.19013}{ \epsilon 
   \sqrt{1- \epsilon }-3.55539 \sqrt{1- \epsilon }-2.35495 \epsilon +3.59101}\,.
\end{equation}
Throughout the text, we will abuse notation and denote all operators by their GFF proxies: the leading $\mathbb{Z}_2$ odd operator is $\phi$, the leading $\mathbb{Z}_2$ even operator is $\phi^2$, the subleading $\mathbb{Z}_2$ odd operator is $\phi^3$ and the subleading $\mathbb{Z}_2$ even operator is $\phi^4$.

\subsection{Lattice model and MC algorithm} \label{ssec:lattice}

We realize the LRI on a chain of $L$ Ising spins, $\sigma_i=\pm1$, with periodic boundary conditions $\sigma_{i+L}=\sigma_i$ and zero magnetic field. Setting the lattice spacing to one, we define the dimensionless finite-volume Hamiltonian as
\begin{equation}
    \beta H_{\mathrm{LRI}}
    =-J_0\sum_{i<j}
    \frac{\sigma_i\sigma_j}{d(i,j)^{1+s}},
    \qquad
    d(i,j)=\min\left(|i-j|,L-|i-j|\right),
\end{equation}
where $J_0$ is the coupling coefficient of each unordered pair. We use the minimum-image convention rather than summing the interaction over periodic images, since minimizing finite-volume corrections through the choice of interaction kernel is not essential for our purposes. In fact having some clearly measurable finite volume effect will give us access to some interesting observables. 

\medskip

A major computational difficulty of long-range lattice models is that each spin interacts with every other spin, so a naive implementation of the Swendsen-Wang cluster algorithm \cite{PhysRevLett.58.86} requires $O(L^2)$ operations per sweep. Luijten and Bl\"ote~\cite{1995IJMPC...6..359L} introduced an efficient alternative based on sampling successive candidate bonds from cumulative bond probabilities, reducing the computational cost to $O(L\log L)$ per sweep. The Fukui-Todo algorithm~\cite{FUKUI20092629} instead uses Poisson sampling and alias tables to obtain an exact cluster update with effectively linear computational cost\footnote{More precisely, using union-find with path compression and union by size gives an expected complexity of $O(L\alpha(L))$, where $\alpha$ is the inverse Ackermann function. Since $\alpha(L)$ grows extremely slowly, the algorithm is effectively linear.}.

For a pair of sites $(i,j)$, let $\mathcal J_{ij}=J_0/d(i,j)^{1+s}$ denote the dimensionless coupling. In the Swendsen-Wang algorithm, a bond between aligned spins is activated with probability
\begin{equation}
    p_{ij}=1-e^{-2\mathcal J_{ij}}\,.
\end{equation}
Instead of testing this probability independently for all $L(L-1)/2$ pairs, the Fukui-Todo algorithm first samples the total number of candidate bonds, $K$, from a Poisson distribution,
\begin{equation}
    K\sim\operatorname{Poisson}(\Lambda),
    \qquad
    \Lambda=2\sum_{i<j}\mathcal J_{ij}\,.
\end{equation}
Each candidate is then assigned to a pair $(i,j)$ with probability
\begin{equation}
    P(i,j)=\frac{\mathcal J_{ij}}
    {\displaystyle\sum_{k<l}\mathcal J_{kl}}\,.
\end{equation}
A bond is activated if the two spins are aligned. Multiple candidates assigned to the same pair are equivalent to a single activated bond.

The computational advantage follows from the summability of the long-range interaction. For the periodic chain considered here,
\begin{equation}
    \Lambda
    =2\sum_{i<j}\mathcal J_{ij} =LJ_0\sum_{r=1}^{L-1}
    \frac{1}{d(0,r)^{1+s}}
    <L J_0 2\zeta(1+s) =O(L)\,,
    \qquad s>0\,.
\end{equation}
Thus, the expected number of candidate bonds grows linearly with the system size, despite the quadratic number of interacting pairs. Translational invariance permits constant-time sampling of each candidate: a site $i$ is selected uniformly, and a displacement $r\in\{1,\ldots,L-1\}$ is drawn with probability proportional to $d(0,r)^{-(1+s)}$, using a precomputed alias table \cite{92917}. The second site is $j=(i+r)\bmod L$. The alias table requires $O(L)$ preprocessing time and memory, but allows each displacement to be sampled in $O(1)$ time, compared with $O(\log L)$ using binary search in the cumulative distribution. This reduces the expected sampling cost per sweep of the original implementation, \cite{PhysRevB.56.8945}, from $O(L\log L)$ to $O(L)$. Clusters are constructed using a disjoint-set union (union-find) data structure with path compression and union by size, giving an overall effectively linear complexity.

The validity of this construction follows from the properties of Poisson distributions. The numbers of candidates assigned to distinct pairs are independent Poisson random variables with means $2\mathcal J_{ij}$. Consequently, the probability that an aligned pair receives at least one candidate is
\begin{equation}
    P(K_{ij}\geq 1)
    =1-P(K_{ij}=0)
    =1-e^{-2\mathcal J_{ij}}\,.
\end{equation}
Since the candidate counts are independent, this construction reproduces the full joint Swendsen--Wang bond distribution. Once the connected clusters are identified, each is independently assigned one of the two spin orientations with equal probability. The resulting Markov transition kernel is therefore identical to that of the Swendsen-Wang algorithm. The algorithm thus enables efficient cluster updates for large systems while retaining the full long-range interaction without approximation.

\section{Critical temperature and deviations to criticality}\label{sec:MC}

In this section, we will determine the critical temperature for several values of the parameter $s$ and consider some effects of detuning.
First, in Section \ref{ssec:binder} we determine the critical coupling \(J_{0,c}\) using standard finite-size scaling methods, based on the Binder cumulant and the crossings of the rescaled spin propagator. We complement these estimates in Section \ref{ssec:critfromprotect}  by exploiting the protected scaling dimension of the spin field, \(\Delta_\phi=(1-s)/2\), which fixes the critical behavior of its two-point function, \(G(r)\sim r^{-2\Delta_\phi}\). This comparison provides an independent check of the critical coupling and probes residual finite-size corrections.

In Section \ref{ssec:offcrit}  we examine the off-critical behavior of the spin correlator, which, unlike in short-range models, retains power-law decay in the disordered phase. We characterize the crossover between the critical and off-critical power laws using a phenomenological propagator that reproduces the qualitative behavior, although systematic deviations remain.

\subsection{Binder and spin propagator}
\label{ssec:binder}
The study of the Binder cumulant for the magnetization $M$ \cite{Binder:1981sa}, defined as
\begin{equation}
\label{binder}
    U_L = 1-\frac{\langle M^4\rangle}
                    {3\langle M^2\rangle^2},
\end{equation}
is the standard method to determine critical temperatures. In the disordered phase, the magnetization distribution becomes Gaussian, so that $U_L\to 0$. For $0<s<1/2$, the quartic coupling is dangerously irrelevant for the zero-momentum mode \cite{PhysRevLett.116.115701}, modifying the finite-size scaling of the Binder cumulant and complicating the identification of crossings between different volumes. We therefore define a pseudocritical coupling by the fixed-threshold condition $U_L=0.1$, whose finite-size drift is analyzed below. The choice of threshold is arbitrary, but its location approaches the critical coupling as $L\to\infty$.

As an alternative estimator, we consider the momentum-space spin propagator,
\begin{equation}
    \widetilde G_L(k)
    = \sum_{r=0}^{L-1} e^{ikr}G_L(r)
    = \frac{1}{L}\sum_{x,y=0}^{L-1}
      e^{ik(x-y)}\langle\sigma_x\sigma_y\rangle,
\end{equation}
where $k=2\pi n/L$ and we assume periodic boundary conditions. Note that we use tildes to denote discrete Fourier transformed quantities throughout the text. We focus on the lowest nonzero momentum, $k_m=2\pi/L$, which eliminates the disconnected contribution and avoids the modified finite-size scaling associated with the zero-momentum mode in the Gaussian regime.

At criticality, the infinite-volume propagator behaves as
\begin{equation}
    \widetilde G(k)\sim |k|^{2\Delta_\phi-1},
    \qquad
    \Delta_\phi=\frac{1-s}{2}.
\end{equation}
For a finite system, the corresponding scaling form is
\begin{equation}
    \widetilde G_L(k_n;J_0)
    = L^{1-2\Delta_\phi}
      \mathcal F(2\pi n,tL^{1/\nu})+\cdots,
\end{equation}
where $t=(J_{0,c}-J_0)/J_{0,c}$, $\nu$ is the correlation-length exponent, and the ellipsis denotes corrections to scaling. Define
\begin{equation}
    \label{eq:definitionF}
    F_L(J_0)
    \equiv L^{2\Delta_\phi-1}
    \widetilde G_L(2\pi/L;J_0)
    =\mathcal F(2\pi,tL^{1/\nu})+\cdots,
\end{equation}
which, at the critical coupling, $F_L$ approaches a volume-independent value. $J_{0,c}$ is then estimated from the intersections of $F_L$ for successive system sizes.

The scaling variable $tL^{1/\nu}$ also determines the width of the critical region,
\begin{equation}
    |\Delta t|\sim L^{-1/\nu}.
\end{equation}
 This scale characterizes the finite-size crossover, and it is a proxy for the systematic uncertainty in the determination of $J_{0,c}$. In the Gaussian regime, $\nu=1/s$, yielding
\begin{equation}
    |\Delta t|\sim L^{-s}.
\end{equation}

\begin{figure}
    \centering
    \includegraphics[width=0.9\columnwidth]
    {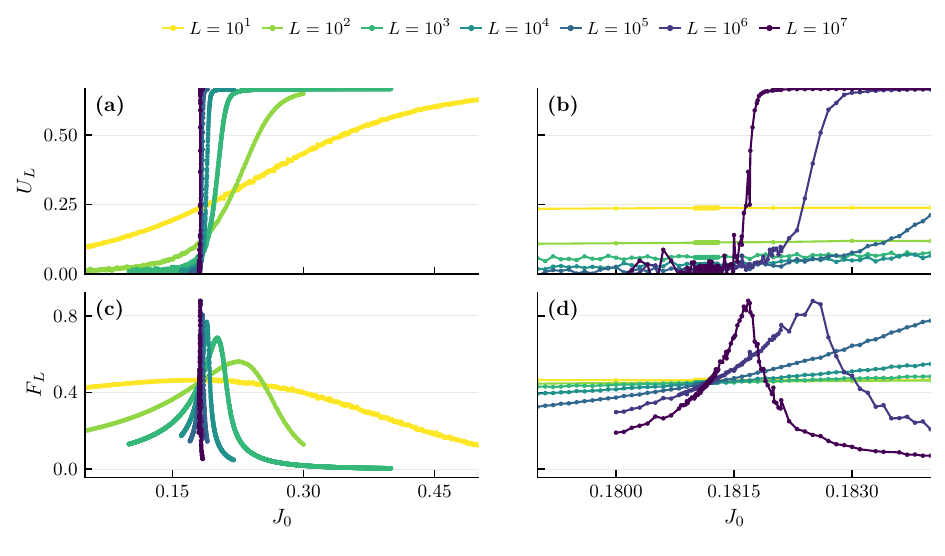}
    \caption{Finite-size behavior at $s=0.4$. (a) Binder cumulant $U_L$; (b) rescaled lowest-momentum correlator $F_L=L^{-s}\widetilde G(2\pi/L)$.  Lines join measured points, to guide the eye. Left) Wide range; Right) Zoom-in on the critical region}
    \label{fig:binder vs F}
\end{figure}

We also considered the ratio
\begin{equation}
    Q_L(J_0)=
    \frac{\widetilde G_L(k_m;J_0)}
         {\widetilde G_L(2k_m;J_0)},
\end{equation}
which approaches unity in the disordered phase and a nontrivial constant at criticality, without requiring prior knowledge of $\Delta_\phi$. However, this observable exhibited larger statistical uncertainties than $F_L$. We therefore exploit the protected scaling dimension $\Delta_\phi=(1-s)/2$ and focus on $F_L$.

\begin{figure}
    \centering
    \includegraphics[width=0.9\columnwidth]
    {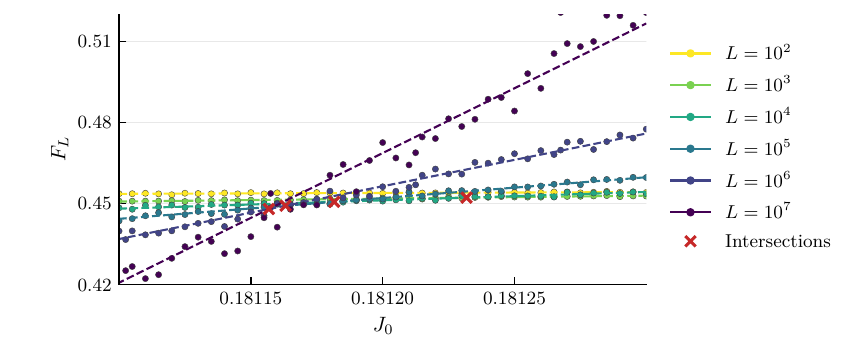}
    \caption{Further zoom of fig.~\ref{fig:binder vs F} in the critical region. Red crosses mark intersections of fits at consecutive lattice sizes. Dashed curves are linear fits over $0.1810\leq J_0\leq0.1813$. }
    \label{fig:F intersection}
\end{figure}

We first compare the two estimators at $s=0.4$, where the infrared fixed point is Gaussian. Left panel of figure~\ref{fig:binder vs F} shows the Binder cumulant and the normalized propagator for different system sizes. The region near the critical point is zoomed in on the right panel of fig.~\ref{fig:binder vs F}. At this value of $s$, the intersections of $F_L$, shown as red crosses in in fig.~\ref{fig:F intersection}, exhibit substantially smaller finite-size drift than the Binder estimator, becoming sharper with increasing $L$.

\begin{figure}
    \centering
    \includegraphics[width=0.9\columnwidth]
    {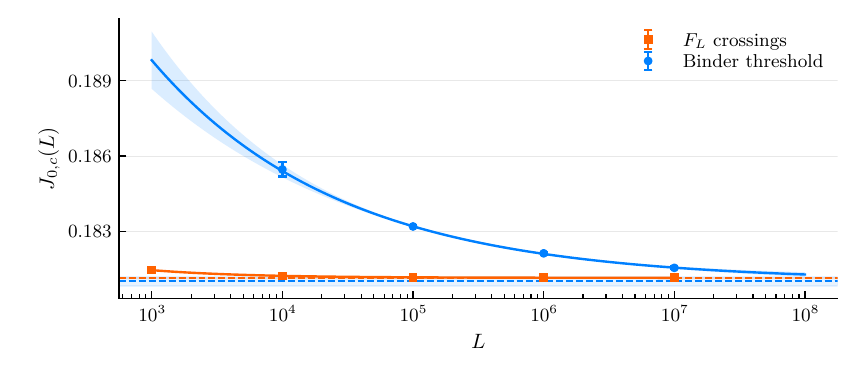}
    \caption{Finite-size critical-coupling estimates at $s=0.4$ from the Binder threshold (blue circles) and consecutive-size crossings of $F_L$ (orange squares). The x-axis coordinate is the larger size in each pair. Solid curves fit $J_{0,c}(L)=J_{0,c}+A L^{-p}$ and dashed lines represent the extrapolated values.  One-standard-error uncertainties are represented by bands.}
    \label{fig:critical js}
    \centering
    \includegraphics[width=0.9\columnwidth]
    {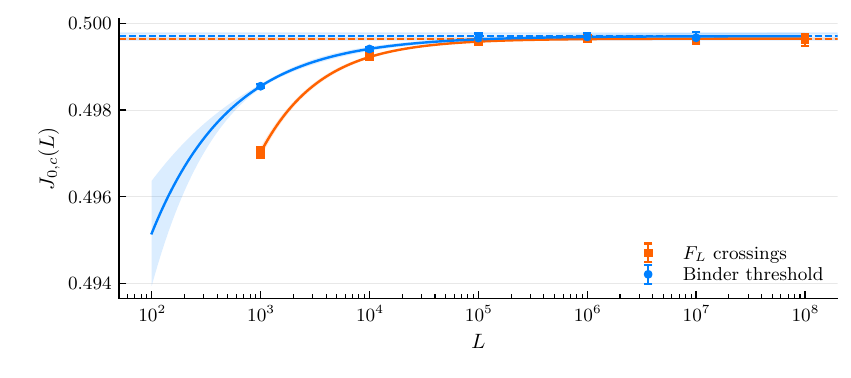}
    \caption{Same as fig.~\ref{fig:critical js}, but for $s=0.9$, where the fixed-point in non-Gaussian and the Binder cumulant provides a more stable estimate of the critical coupling.}
    \label{fig:critical js s=00003D0.8}
     \centering
    \begin{minipage}[c]{0.39\textwidth}
        \centering
        \includegraphics[width=\linewidth]{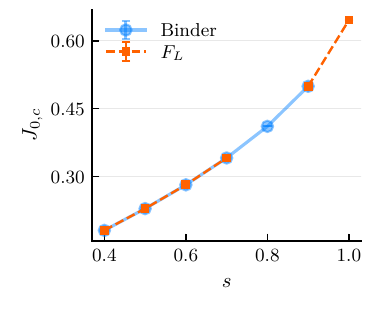}\\[0.8em]
    \end{minipage}
    \hfill
    \begin{minipage}[c]{0.59\textwidth}
        \centering
        \resizebox{\linewidth}{!}{%

\begin{tabular}{ccccccc}
    \hline
    $s$ &
    $J_{0,c}^{\mathrm B}$ &
    $\alpha_{\mathrm B}$ &
    $J_{0,c}^{\mathrm F}$ &
    $\alpha_{\mathrm F}$ &
    \cite{PhysRevE.63.037101} &
    \cite{MARTINEZHERRERA2022127136} \\
    \hline
    0.4 & 0.18102(18)  & 0.30(4) & 0.181158(2)   & 0.59(3) & ---          & 0.213(27) \\
    0.5 & 0.22917(12)  & 0.4(3)  & 0.2291528(28) & 0.54(6) & ---          & 0.270(37) \\
    0.6 & 0.281802(11) & 0.73(2) & 0.281804(4)  & ---     & 0.28249(48)  & 0.333(44) \\
    0.7 & 0.341243(10) & 0.81(2) & 0.341264(9)  & 0.63(6) & 0.34176(70)  & 0.385(44) \\
    0.8 & 0.411109(2)  & 0.83(4) & ---           & ---     & 0.41135(68)  & 0.435(57) \\
    0.9 & 0.49971(8)   & 0.6(1)  & 0.49965(2)    & 0.79(5) & 0.49950(50)  & 0.500(50) \\
    1.0 & ---           & ---     & 0.6461(4)     & 0.54(2) & ---          & 0.556(62) \\
    \hline
\end{tabular}

        }
    \end{minipage}
    \caption{Critical couplings and  exponents from the Binder and rescaled-propagator crossings. The table shows the numbers plotted. Missing entries indicate unavailable or unreliable estimates. The values compare well with \cite{PhysRevE.63.037101}. The estimates of Mart\'inez-Herrera, Rodr\'iguez-L\'opez, and Sol\'is~\cite{MARTINEZHERRERA2022127136}, obtained through finite-range transfer-matrix extrapolation, provide an independent consistency check, although with substantially larger uncertainties.}
    \label{fig:J0c as a function of the temperature.}
\end{figure}

For a quantitative comparison, we define $J_{0,c}^{\mathrm B}(L)$ by the condition $U_L=0.1$ and $J_{0,c}^{\mathrm F}(L)$ by the intersection of $F_L$ at successive system sizes. We extrapolate both estimates using
\begin{equation}
    J_{0,c}^{X}(L)
    =J_{0,c}^{X}+A_XL^{-\alpha_X},
    \qquad X\in\{\mathrm B,\mathrm F\}.
\end{equation}
For $s=0.4$, the fits give
\begin{align}
    J_{0,c}^{\mathrm B}(L)
    &=0.18102(18)+0.071(28)L^{-0.30(4)},\\
    J_{0,c}^{\mathrm F}(L)
    &=0.181158(2)+0.018(4)L^{-0.59(3)}.
\end{align}
The extrapolated critical couplings agree within their quoted uncertainties. The propagator estimator exhibits a faster decay of the fitted finite-size corrections and a substantially smaller extrapolation uncertainty, shown in the table in fig.~\ref{fig:critical js}.

The relative performance of the estimators depends on $s$. At $s=0.9$, within the interacting regime, the Binder estimator shows smaller finite-size deviations over the range of volumes considered, as illustrated in fig.~\ref{fig:critical js s=00003D0.8}. Results for other values of $s$ are presented in app.~\ref{app:all critical couplings finite volume dependence}. The critical couplings and fitted exponents are summarized in fig.~\ref{fig:J0c as a function of the temperature.}. In general, the Binder and the $F_L$ estimates are compatible. 

The fitted exponents $\alpha_X$ characterize the finite-size corrections over the range of volumes considered. Since several correction terms may contribute, these effective exponents cannot be unambiguously identified with universal correction-to-scaling exponents. At the marginal values $s=1/2$ and $s=1$, logarithmic corrections may further complicate the finite-size scaling analysis \cite{hym1-lmwt}.

\subsection{Criticality from the protected scaling dimension}
\label{ssec:critfromprotect}

\begin{figure}
    \centering
    \includegraphics[width=\linewidth]
    {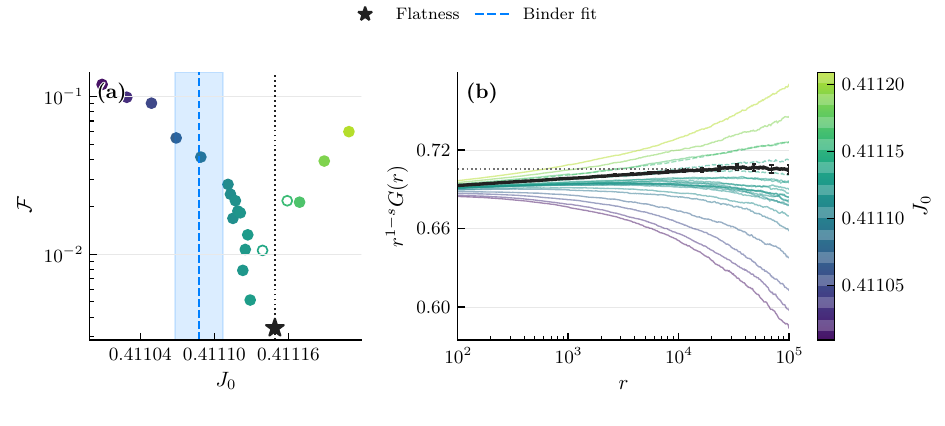}
    \caption{Critical-coupling selection at $s=0.8$ and $L=10^7$. (a) Flatness scores for 22 couplings around $J_{0,c}^{\mathrm B}$. The black star and dotted line mark the coupling selected by flatness, $J_{0,c}^{\mathcal F}=0.411149$. The blue dashed line and band show the Binder estimate $J_{0,c}^{\mathrm B}=0.411087(20)$ and its uncertainty. (b) Compensated correlators over $100\leq r<10^5$. The selected curve is black; the blue dash-dotted curve corresponds to the earlier simulation coupling $J_0=0.411109$.}
    \label{fig:flatness score}
\end{figure}
The scaling dimension of the spin field, $\Delta_\phi=(1-s)/2$, is protected by the nonlocal kinetic term. At criticality, the infinite-volume two-point function therefore behaves as $G(r)\propto r^{-2\Delta_\phi}$. We use this prediction as an additional check of the critical coupling obtained from the finite-size scaling analysis. For a given coupling $J_0$, we define the rescaled correlator
\begin{equation}
    y_j(J_0)=r_j^{2\Delta_\phi}G_L(r_j;J_0).
\end{equation} 
We consider only $r\in[100,L/100]$ to mitigate finite volume and finite lattice spacing effects.

To quantify deviations from the predicted behavior, we introduce a heuristic measure of the flatness of the data, $\mathcal{F}$,
\begin{equation}
\begin{split}
    m &= \operatorname{median}(y_j), \qquad
    S = \max\left(|m|,\frac{1}{n}\sum_j |y_j|\right),\\
    d &= \frac{
        \operatorname{median}(y_{\mathrm{last\ third}})
        -\operatorname{median}(y_{\mathrm{first\ third}})
    }{S},\\
    \mathcal F &=
    \sqrt{\frac{1}{n}\sum_j
    \left(\frac{y_j-m}{S}\right)^2}+|d|.
\end{split}
\label{eq:lri-num-flatness}
\end{equation}
The median $m$ provides an estimate of the overall amplitude, while $S$ makes the score insensitive to its normalization and avoids an artificially small denominator when the median is close to zero. The first term in $\mathcal{F}$ measures the relative deviations from a constant, while the second explicitly penalizes systematic drift between short and long distances (e.g. coherent deviations from the expected power law). 

We identify the coupling that minimizes $\mathcal F$ and visually inspect the rescaled correlators at the selected and neighboring couplings, as illustrated in fig.~\ref{fig:flatness score} for $s=0.8$. The selected value, $J_{0,c}=0.411149$, approximately minimizes the distance dependence of the rescaled correlator, while nearby couplings exhibit systematic upward or downward drifts. 

At $s=0.8$, where only the Binder cumulant is available, its estimate lies significantly bellow the one selected by the flatness criteria, and the corresponding rescaled correlator bends downward at large distances, consistent with detuning.  For us, this tension is only relevant in so far that reflectes in the determination of $\Delta_{\phi^2}$. In fig.~\ref{fig:energy-coupling-comparison} in subsec.~\ref{ssec:correlation-functions}, we consider simulations at both values of the coupling for selected $s$ and show that, at the current signal-to-noise ratio, they yield indistinguishable predictions for $\Delta_{\phi^2}$. 

\subsection{Off-critical physics}\label{ssec:offcrit}

The curvature of the rescaled spin correlators in fig.~\ref{fig:flatness score} has a direct interpretation in terms of the off-critical infrared behavior. In contrast to short-range systems, long-range interactions produce power-law correlations even away from the critical point; this is particularly easy to see in the regime $0<s<1/2$ as discussed in Section \ref{ssec:QFT}. 

The long-range interaction $J(r)=J_0/r^{1+s}$ determines the momentum dependence of the quadratic part of the effective action. In Fourier space, 
\begin{equation}
    \widetilde J(0)-\widetilde J(k)
    =2J_0\sum_{r=1}^{\infty}
    \frac{1-\cos(kr)}{r^{1+s}}
    \simeq 2J_0 |k|^s
    \int_0^\infty du\,\frac{1-\cos u}{u^{1+s}},
\end{equation}
where the integral converges for $0<s<2$. Thus, unlike short-range interactions, which generate an analytic $k^2$ dependence, the long-range interaction produces a nonanalytic $|k|^s$ term. This term controls the long-wavelength fluctuations, giving $\widetilde G(k)\propto |k|^{-s}$ at criticality. Away from criticality, the inverse propagator also acquires a nonzero contribution at $k=0$, leading to the infrared form
\begin{equation}
    \widetilde G(k,t)^{-1}
    \simeq \chi(t)^{-1}+c|k|^s,
\end{equation}
where $\chi(t)=\widetilde G(0,t)$ is the susceptibility.
At criticality, $\chi^{-1}\to0$, giving the protected power $G(r;0)\propto r^{-(1-s)}$. For $t>0$, the expansion around $k=0$ gives $\widetilde G(k;t)=\chi-A_s\chi^2|k|^s+\cdots$. The nonanalytic term implies $G(r;t)\propto r^{-(1+s)}$ at asymptotically large separations, rather than exponential decay. Thus, the crossover length $\xi\sim t^{-\nu}$ \footnote{The power-law thermal scaling used above applies for $0<s<1$. The endpoint $s=1$ corresponds to the inverse-square Ising model, which exhibits a Thouless transition with Kosterlitz-Thouless-like critical behavior \cite{PhysRevLett.86.5305}. The crossover length diverges with an essential singularity, $\xi\sim\exp(B/\sqrt{t})$, rather than a power law, and the critical spin correlator approaches its nonzero
large-distance plateau with a logarithmically decaying correction. Nevertheless, the off-critical algebraic tail $G(r)\propto r^{-2}$ remains valid in the disordered phase.},
rather than an exponential screening length, characterizes the crossover between two algebraic regimes:
\begin{equation}
    G(r;t)=r^{-(1-s)}f(r/\xi),\qquad
    f(z)\sim
    \begin{cases}
        \mathrm{const}, & z\ll1,\\
        z^{-2s}, & z\gg1.
    \end{cases}
    \label{eq:offcritical-crossover}
\end{equation}

In particular, $r^{1-s}G(r;t)$ bends downward at $r\gtrsim\xi$ on the disordered side, as observed in fig.~\ref{fig:flatness score}. Conversely, above the critical coupling, the system develops long-range magnetic order, so the compensated correlator bends upward at large distances.

To describe the crossover observed in the Monte Carlo correlators, we construct a phenomenological propagator that incorporates the infrared structure of the long-range interaction while retaining the finite-volume lattice geometry.

The leading nonanalytic momentum dependence is determined by the interaction kernel, $J(r)\propto r^{-(1+s)}$. On the periodic lattice, its counterpart is 
\begin{equation}
    D_{s,L}(k)
    =\sum_{u=1}^{L-1}
    \frac{1-\cos(ku)}{d(0,u)^{1+s}},
\end{equation}
where the overall coupling $J_0$ has been omitted. For even $L$  and $k=2\pi n/L$, this becomes
\begin{equation}
    D_{s,L}(k)=
    2\sum_{u=1}^{L/2-1}
    \frac{1-\cos(ku)}{u^{1+s}}
    +\frac{1-\cos(kL/2)}{(L/2)^{1+s}}.
    \label{eq:crossover-ring-kernel}
\end{equation}

\begin{figure}
    \centering
    \includegraphics[width=\linewidth]
    {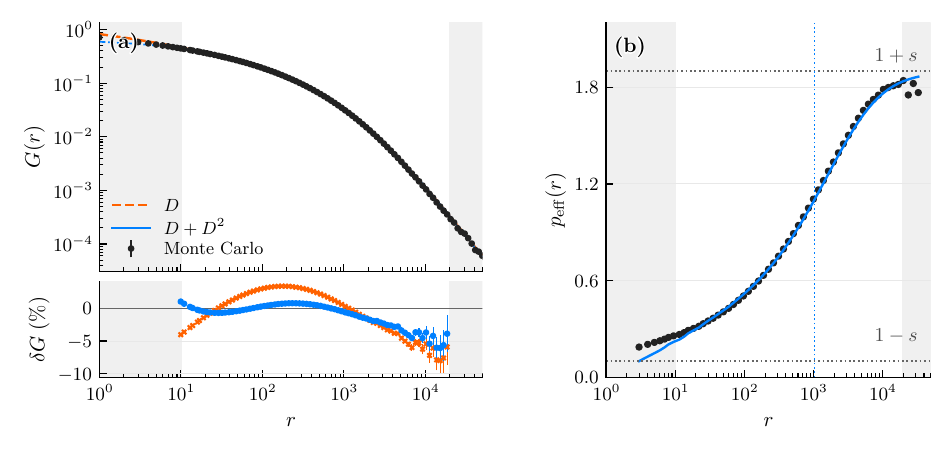}
    \caption{Off-critical spin correlations at $s=0.9$, $J_0=0.47$, and $L=10^7$. (a) Correlator and finite-ring fractional-kernel fits:  baseline $D$ ($b_{2s}=0$) and the $D+D^2$ model. The lower panel gives $\delta G=100(G/G_{\rm fit}-1)$. Fits use all raw separations $10\leq r\leq2\times10^4$, where the upper bound is due to the statistical noise. Shaded distances lie outside that interval. The  measured correlator fixes $\chi=323(2)$. (b) Smoothed local effective exponents, $-\mathrm{d}\log G(r)/\mathrm{d}\log r$, computed from the data and model. The horizontal dotted lines indicate $1-s$ and $1+s$, while the vertical dotted line marks $\xi_{\rm cont}=(a c_s)^{1/s}\simeq1037$.}
    \label{fig:offcritical-crossover}
\end{figure}

In the disordered phase, the infrared inverse propagator has the form
\begin{equation}
    \widetilde G(k)
    =\dfrac{1}{\chi^{-1}+A|k|^s+\cdots}\,.
\end{equation}
Motivated by this expression, we use the fitting ansatz
\begin{equation}
    \widetilde G_L^{\mathrm{fit}}(k)
    =\frac{\chi}
    {1+aD_{s,L}(k)+b_{2s}D_{s,L}(k)^2},
    \label{eq:crossover-fractional-fit}
\end{equation}
where $a>0$ determines the crossover length $\xi_{\mathrm{fit}}=(a c_s)^{1/s}$. A critical-like regime, $G(r)\propto r^{-(1-s)}$, arises for $1\ll r\ll\xi_{\mathrm{fit}}$, when $aD$ dominates the denominator, while for $r\gg\xi_{\mathrm{fit}}$ it approaches the off-critical algebraic tail $G(r)\propto r^{-(1+s)}$. The term proportional to $b_{2s}$, just allows the shape of this crossover to vary, without changing the leading long-distance asymptotics. The real-space correlator through the inverse discrete Fourier transform,
\begin{equation}
    G_L^{\mathrm{fit}}(r)
    =\frac1L\sum_{n=0}^{L-1}
    e^{2\pi i nr/L}
    \widetilde G_L^{\mathrm{fit}}
    \left(\frac{2\pi n}{L}\right),
\end{equation}
which is compared directly with the Monte Carlo data. 

The susceptibility is fixed from the same Monte Carlo ensemble
\begin{equation}
    \chi=\frac{\langle M^2\rangle-\langle M\rangle^2}{L}=\sum_rG(r),
    \qquad M=\sum_i\sigma_i,
\end{equation}
so that only $a$ and $b_{2s}$ are fitted. Since $D_{s,L}(0)=0$, the ansatz automatically reproduces the measured zero-momentum correlator.

Figure~\ref{fig:offcritical-crossover} shows the fit at $s=0.9$. The quadratic-correction ansatz captures the overall shape of the off-critical correlator and its approach towards the predicted $r^{-(1+s)}$ tail, although systematic deviations remain. 

For $0<s<2$, the infinite-volume kernel satisfies $D_{s,\infty}(k)\sim c_s|k|^s$ as $k\to0$, with $c_s=\pi/[\Gamma(1+s)\sin(\pi s/2)]$. The linear term in the fitted denominator defines the crossover scale $\xi_{\mathrm{fit}}=(a c_s)^{1/s}$.
In the disordered phase, where $\chi$ is finite, the small-momentum expansion of the fit gives
\begin{equation}
    G^{\mathrm{fit}}(r)\sim\frac{\chi a}{r^{1+s}},
    \label{eq:crossover-fit-tail}
\end{equation}
in the bulk large-distance limit. On a finite ring, this requires $1\ll r\ll L$.

For the infinite-volume ferromagnetic Ising model with these power-law couplings, in the disordered phase $J_0<J_{0,c}$ and at zero magnetic field, correlation asymptotics imply~\cite{Aoun2021}
\begin{equation}
    G(r)\sim \chi^2 J_0 r^{-(1+s)},
    \qquad r\to\infty.
    \label{eq:offcritical-tail-amplitude}
\end{equation}
Matching the asymptotic amplitudes, gives
\begin{equation}
    a=\chi J_0.
\end{equation}
The quadratic-correction fit yields $a/(\chi J_0)\simeq1.032$. Since $a$ is extracted from a phenomenological crossover fit with residual systematic deviations, the ratio should not be interpreted as an independent validation of the asymptotic amplitude to $3.2\%$ accuracy.

\begin{figure}
    \centering
    \includegraphics[width=\linewidth]{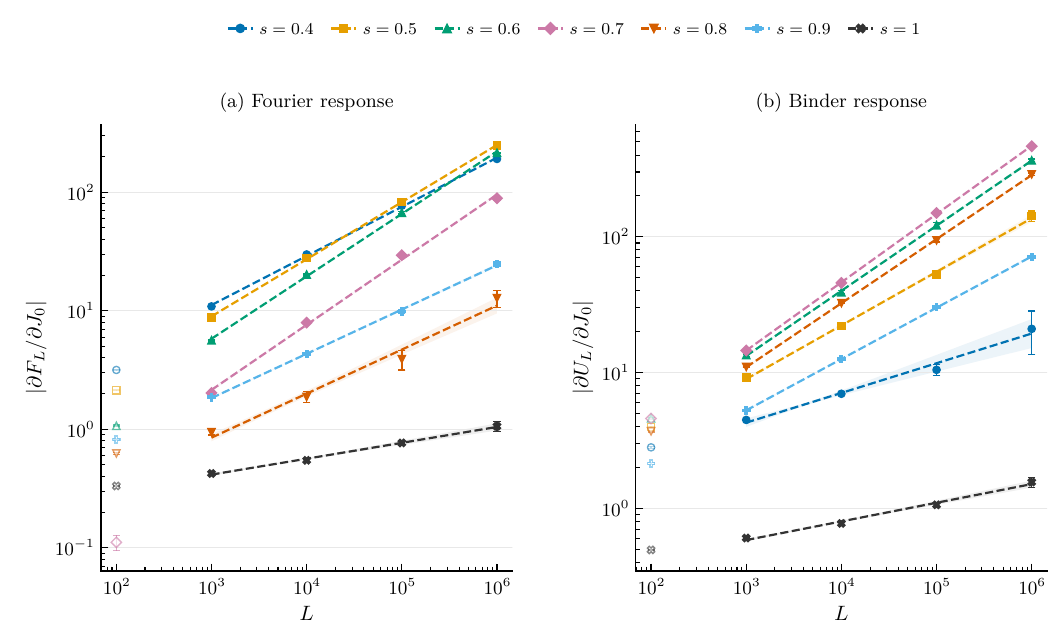}
    \caption{Thermal responses from covariance at the critical points.  The $F_L$ response is negative and the Binder response positive.}
    \label{fig:block-thermal-response}

    \vspace{1em}

    \begin{minipage}{\linewidth}
    \centering
    \small
    \setlength{\tabcolsep}{4.5pt}
    \renewcommand{\arraystretch}{1.25}
    \begin{tabular}{cccccc}
    \toprule
    $s$ & $J_0$ & \multicolumn{2}{c}{$\Delta_{\phi^2}^{F}$}
    & \multicolumn{2}{c}{$\Delta_{\phi^2}^{U}$} \\
    \cmidrule(lr){3-4}\cmidrule(lr){5-6}
    & & $L_{\min}=10^3$ & $L_{\min}=10^4$
    & $L_{\min}=10^3$ & $L_{\min}=10^4$ \\
    \midrule
    $0.4^{\mathrm{G}}$ & 0.181158 & $0.5855(12)^{\dagger}$ & $0.5970(20)^{\dagger}$ & $0.781(45)$ & $0.761(76)$ \\
    $0.5^{\mathrm{M}}$ & 0.2291528 & $0.5179(12)^{\dagger}$ & $0.5258(20)$ & $0.605(12)$ & $0.595(19)$ \\
    $0.6$ & 0.281802 & $0.4731(22)^{\dagger}$ & $0.4866(27)$ & $0.5204(47)$ & $0.5141(81)$ \\
    $0.7$ & 0.341243 & $0.4506(65)^{\dagger}$ & $0.4755(93)^{\dagger}$ & $0.4972(36)$ & $0.4963(64)$ \\
    $0.8$ & 0.411109 & $0.628(27)^{\ddagger}$ & $0.583(47)$ & $0.5281(32)$ & $0.5257(50)$ \\
    $0.9$ & 0.49971 & $0.6264(61)^{\ddagger}$ & $0.6194(99)^{\ddagger}$ & $0.6221(43)^{\ddagger}$ & $0.6234(66)^{\ddagger}$ \\
    $1^{\mathrm{KT}}$ & 0.6461 & $0.866(13)$ & $0.857(22)$ & $0.863(11)$ & $0.848(18)$ \\
    \bottomrule
    \end{tabular}

    \captionsetup{type=table}
    \caption{Log-power fits of  $|\partial X_L/\partial J_0|    =A_X(L/10^4)^{1-\Delta_{\phi^2}^{\mathrm{eff}}}$    of the data in Fig.~\ref{fig:block-thermal-response},  where $L_{\min}$ denotes the smallest size included in the fit. Critical-coupling and scaling-correction uncertainties are excluded. $\dagger$: the simple power law fits poorly; $\ddagger$: errors possibly underestimated. G: Gaussian; M: marginal point; KT: KT-like transition.}
    \label{tab:direct-response-final-L1000}
    \end{minipage}
\end{figure}

\section{Thermal response of the spin correlator} \label{ssec:thermal-response}

The coupling dependence of the momentum space spin propagator and Binder cumulant provide indirect access to the observable $\Delta_{\phi^2}$. Therefore in this section we obtain estimates for this quantity which will be complemented by direct measurements of the scaling dimension via correlation functions in Section \ref{sec:results}.

In the interacting regime, $1/2<s<1$, the finite-size scaling form of eq.~\eqref{eq:definitionF} gives
\begin{equation}
 \left.\frac{\partial F_L}{\partial J_0}\right|_{J_{0,c}}
 = A_F L^{y_t}\left(1+B_F L^{-\omega_F}+\cdots\right),
 \qquad y_t=\frac1\nu=1-\Delta_{\phi^2}.
 \label{eq:thermal-response-slope}
\end{equation}
While the protected $\Delta_\phi$ fixes the scale of $G_L(r_j)$, its thermal response probes $\Delta_{\phi^2}$. For sizes $L$ and $bL$, the ratio of the two slopes (as a function of the coupling) approaches $b^{y_t}$\footnote{The same argument applies to the coupling derivative of $U_L$.}. The marginal regimes require their own separate scaling forms.

Naively, the derivative can be evaluated numerically. However, a better estimator comes from the covariance of these quantities.  Writing the dimensionless Hamiltonian as $\beta H_{\mathrm{LRI}}=-J_0 Q$, differentiation of the Boltzmann weight gives
\begin{equation}
    \frac{\partial\langle A\rangle}{\partial J_0}
    =\langle AQ\rangle-\langle A\rangle\langle Q\rangle
\end{equation}
for any observable $A$ without explicit coupling dependence. This allows us to compute the thermal response directly from fluctuations at a single coupling, without numerical differentiation. We apply this identity to the Fourier-mode power entering $F_L$ \eqref{eq:definitionF} and to the magnetization moments defining $U_L$ \eqref{binder}. Table~\ref{tab:direct-response-final-L1000} shows the extracted dimension as a function of the smallest lattices included in the fitz. The data is shown in fig.~\ref{fig:block-thermal-response}. 

At criticality, both derivatives are expected to scale as
\begin{equation}
    \left|\frac{\partial F_L}{\partial J_0}\right|,
    \left|\frac{\partial U_L}{\partial J_0}\right|
    \sim L^{y_t},
    \qquad y_t=\frac{1}{\nu}=1-\Delta_{\phi^2},
\end{equation}
up to sub-leading corrections. At $s=0.9$ and $J_0=0.49971$, both responses exhibit approximately the expected power-law growth, see fig.~\ref{fig:block-thermal-response}, while the values of $\Delta_{\phi^2}$ are shown in tab.~\ref{tab:direct-response-final-L1000}. 

The Binder response is well described by a linear fit in log--log coordinates and yields estimates of $\Delta_{\phi^2}$ that are stable under changes in $L_{\min}$, whereas the Fourier response does not.

\section{Correlation functions}
\label{sec:results}

We now study correlation functions to extract scaling dimensions and operator-product expansion (OPE) data. The relatively small scaling dimensions of the lowest operators lead to slowly decaying correlators, making higher-point functions particularly accessible to Monte Carlo simulations. Their analysis is further simplified in one dimension, where conformal invariance reduces the nontrivial dependence of scalar four-point functions to a single cross ratio.

We first use the protected scaling dimension of the spin field to characterize the choice of local vs non-local lattice operators in Section \ref{subsec:local-nonlocal-spin}, before extracting the leading unprotected even dimension from connected two-point functions in Section \ref{ssec:correlation-functions}  and investigating higher even operators in Section \ref{ssec:unprotected}. Finally, we study three-(Section \ref{ssec:three-point}) and four-point functions (Section \ref{ssec:four-point}) to probe OPE coefficients and conformal structure.

\subsection{Local and nonlocal spin correlators}
\label{subsec:local-nonlocal-spin}

Before extracting $\Delta_{\phi^2}$ from connected energy correlators, we investigate how the nonlocality of lattice observables affects the extracted operator spectrum. The energy density of the long-range model contains a convolution of the spin field with the interaction kernel, closely related to the shadow transform. This makes nonlocal operators natural candidates for extracting continuum operator data. However, nonlocal kernels can introduce subleading power laws that mimic contributions from higher-dimensional operators, complicating their use in a generalized eigenvalue problem (GEVP) basis aimed at resolving heavier states. Conversely, we will see that suitably chosen nonlocal $\mathbb{Z}_2$-even operators can reduce corrections to scaling in two-point functions. These observations guide our choice of operators for the subsequent measurements.

We first isolate the corrections introduced by nonlocality in the spin sector, where the leading scaling dimension $\Delta_\phi=(1-s)/2$ is protected. We compare the local spin $\sigma_i$ with kernel-weighted operators
\begin{equation}
    \Sigma_{K,i}=\sum_{u\ne0}K(u)\sigma_{i+u},
    \qquad
    K(u)=\frac{J_0}{d_L(u)^{1+s_{\mathrm{op}}}},
    \label{eq:nonlocal-spin-definition}
\end{equation}
where $d_L$ is the minimum-image distance and $K(0)=0$ by convention. We consider the matched exponent $s_{\mathrm{op}}=s$, the steeper kernels $s_{\mathrm{op}}=s+1,s+2$, and also a truncated kernel at distance $R$, equal to one. Defining $K_0=\sum_u K(u)$ and $\Phi_{K,i}=\Sigma_{K,i}/K_0$, we compare
\begin{equation}
    G(r)=\langle\sigma_i\sigma_{i+r}\rangle_c,
    \qquad
    C_K(r)=\langle\Phi_{K,i}\Phi_{K,i+r}\rangle_c.
    \label{eq:normalized-spin-correlators}
\end{equation}
Since $\Phi_K$ is linear in the spin, these correlators satisfy the exact relation
\begin{equation}
    \widetilde C_K(k)=q_K(k)^2\widetilde G(k),
    \qquad
    q_K(k)=\frac{\widetilde K(k)}{K_0}.
    \label{eq:spin-filter-identities}
\end{equation}
The normalization ensures $q_K(0)=1$, preserving the leading infrared singularity. Consequently, at criticality,
\begin{equation}
    G(r)\sim C_K(r)\sim A r^{-(1-s)},
    \qquad 1\ll r\ll L.
    \label{eq:filtered-spin-leading-scaling}
\end{equation}

In the bulk small-momentum limit, a power-law kernel with $0<s_{\mathrm{op}}<2$ satisfies
\begin{equation}
    q_K(k)=1-c_K|k|^{s_{\mathrm{op}}}+\cdots,
    \qquad c_K>0,
\end{equation}
whereas finite-range kernels and kernels with $s_{\mathrm{op}}>2$ have leading analytic corrections proportional to $k^2$. The difference from the local correlator is
\begin{equation}
    \widetilde C_K(k)-\widetilde G(k)
    =\left[q_K(k)^2-1\right]\widetilde G(k).
    \label{eq:spin-filter-corrections}
\end{equation}
Combining these corrections with $\widetilde G(k)\sim|k|^{-s}$ gives different subleading powers. The matched case $s_{\mathrm{op}}=s$ is special: the leading product $|k|^s|k|^{-s}$ is a contact term, so the behavior at nonzero separation is controlled by higher-order contributions. In particular, the quadratic term in $q_K(k)^2$ generates a correction proportional to $r^{-(1+s)}$.

For $0<s<1$, the resulting leading kernel-induced relative corrections are generically of order
\begin{equation}
    \frac{C_K(r)-G(r)}{G(r)}
    =
    \begin{cases}
        O(r^{-2s}), & s_{\mathrm{op}}=s,\\
        O(r^{-(1+s)}), & s_{\mathrm{op}}=s+1,
    \end{cases}
    \label{eq:spin-filter-relative-corrections}
\end{equation}
assuming the corresponding leading amplitudes do not vanish.

\begin{figure}
    \centering
    \includegraphics[width=\linewidth]
    {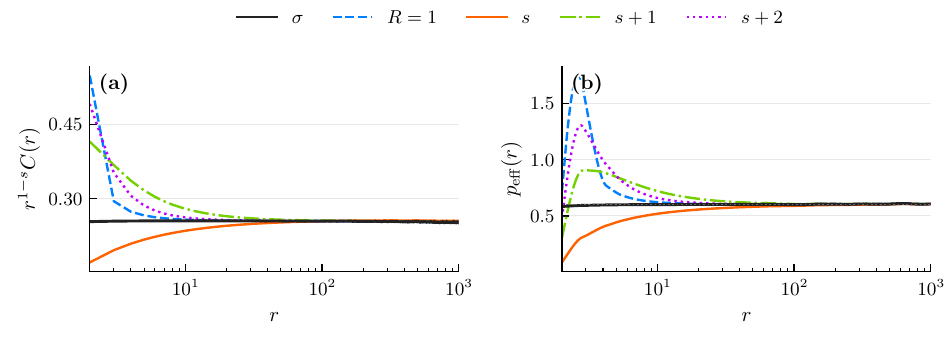}
    \caption{
    Local and kernel-weighted spin autocorrelators at $s=0.4$, $J_0=0.181198$, and $L=2^{18}$. The black curve represents the local spin, while the   remaining curves correspond to normalized filtered spins  $\Phi_K=\Sigma_K/K_0$, using a nearest-neighbor kernel   ($R=1$) and power-law kernels with exponents    $s$, $s+1$, and $s+2$.    (a) Compensated correlators $r^{1-s}C(r)$.   (b) Local effective decay exponents  $p_{\mathrm{eff}}(r)=-d\log C(r)/d\log r$.  All correlators approach the same asymptotic behavior,   but exhibit different finite-distance corrections,  including different signs for the finite-distance corrections.
    }
    \label{fig:local-nonlocal-spin}
\end{figure}

Figure~\ref{fig:local-nonlocal-spin} illustrates these effects at $s=0.4$, where the expected decay exponent is $1-s=0.6$. The local spin correlator is already close to its asymptotic power law over the displayed distances. In contrast, the filtered correlators exhibit substantial finite-distance corrections. The matched-power kernel approaches the asymptotic behavior from below, with an effective exponent that increases towards $1-s$. The nearest-neighbor and steeper power-law kernels approach from above. The nearest-neighbor filter exhibits a particularly large short-distance deviation but converges rapidly at larger separations.

These different apparent decay powers arise entirely from smearing the same spin correlator. This illustrates how nonlocality can complicate the identification of heavier operators in correlation-matrix analyses. Nevertheless, nonlocality need not be detrimental when extracting the leading scaling dimension: as shown in the next subsection, suitable nonlocal energy operators can instead reduce finite-distance corrections.

\subsection{$\mathbb{Z}_2$-even two-point functions}
\label{ssec:correlation-functions}

The leading nontrivial $\mathbb{Z}_2$-even operator, $\phi^2$, has an interaction-dependent scaling dimension. Here, we extract $\Delta_{\phi^2}$ from connected correlators of local and nonlocal energy operators, comparing their corrections to scaling.

We consider three site-centered $\mathbb{Z}_2$-even operators:
\begin{equation}
\begin{aligned}
    E_{\mathrm{loc},i}
    &=J_0\sigma_i(\sigma_{i-1}+\sigma_{i+1}),\\
    E_{p,i}
    &=J_0\sigma_i\sum_{u=1}^{L-1}
    \frac{\sigma_{i+u}}{d_L(u)^{1+p}},
    \qquad p\in\{s,s+1\}.
\end{aligned}
\label{eq:energy-operators}
\end{equation}
The matched-power operator $E_s$ is the natural energy density of the long-range Hamiltonian, $\beta H_{\mathrm{LRI}}=-\frac12\sum_i E_{s,i}$. The steeper kernel $E_{s+1}$ suppresses contributions from distant spins, while $E_{\mathrm{loc}}$ contains only nearest-neighbor pairs.

For each operator, we measure the connected two-point function
\begin{equation}
    C_a(r)=
    \langle E_{a,i}E_{a,i+r}\rangle
    -\langle E_a\rangle^2,
    \qquad a\in\{\mathrm{loc},s,s+1\}.
    \label{eq:energy-connected-correlators}
\end{equation}
Provided the operators have nonzero overlap with $\phi^2$ and kernel-induced contributions are subleading, their critical correlators share the asymptotic behavior
\begin{equation}
    C_a(r)\sim A_a r^{-2\Delta_{\phi^2}},
    \qquad 1\ll r\ll L.
    \label{eq:energy-leading-scaling}
\end{equation}
We characterize their approach to this limit through the local effective dimensions
\begin{equation}
    \Delta_a^{\mathrm{eff}}(r)
    =-\frac12\frac{d\log C_a(r)}{d\log r}.
    \label{eq:energy-effective-dimension}
\end{equation}

\begin{figure}[t]
    \centering
    \includegraphics[trim={0cm 0.8cm 0cm 0cm}, clip, width=\linewidth]{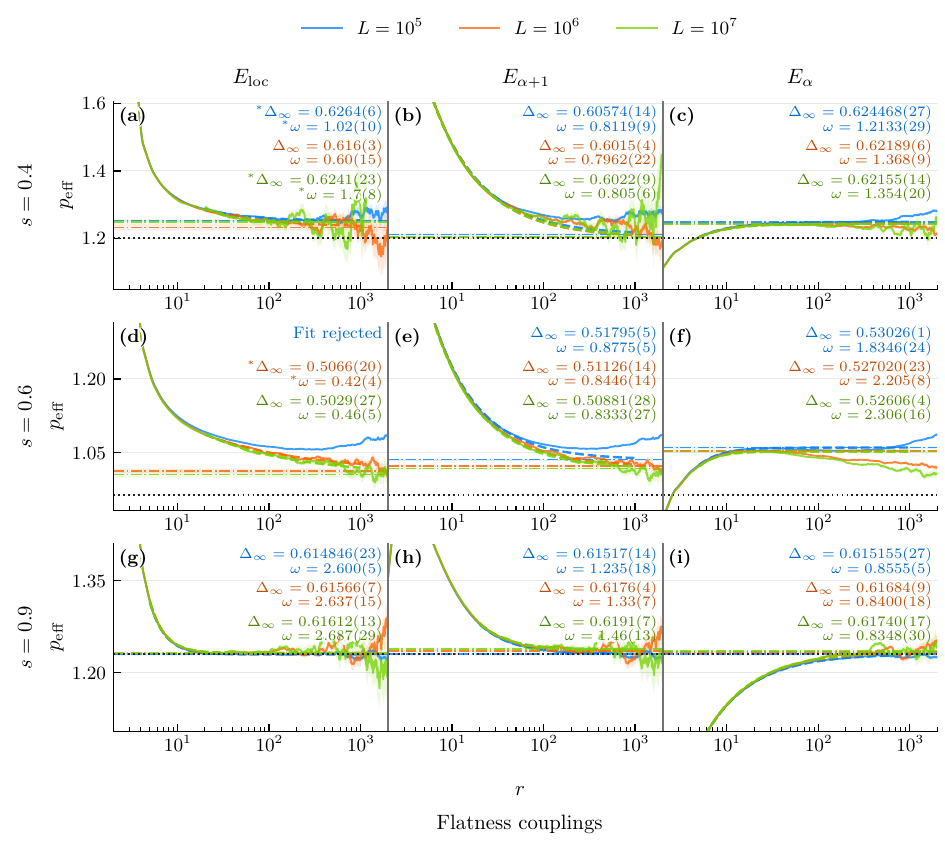}
    \caption{Effective dimensions from connected energy correlators at the spin-flatness couplings $J_0=0.181198,0.281830,0.499740$, for $s=0.4,0.6,0.9$, respectively. Columns correspond to increasing $s$; rows show the local, $s+1$, and matched-power energies. Colors distinguish $L=10^5,10^6,10^7$. Solid curves and bands show the effective dimensions. Dashed curves are fits, and dash-dotted lines with bands indicate their extrapolated limits. Asterisks identify window-sensitive fits. Dotted black lines show the Gaussian/Pad\'e reference. Fits use $5<r<1000$, except for $E_{s+1}$ at $s=0.9$ and the local energy at $s=0.4,0.6$, which use $20<r<1000$. Asterisks mark fits whose parameters are unstable under changes of the fitting window.}
    \label{fig:energy-volume-comparison}
\end{figure}

\begin{figure}[t]
    \centering
    \includegraphics[page=2,trim={0cm 0.8cm 0cm 0cm}, clip,width=\linewidth]{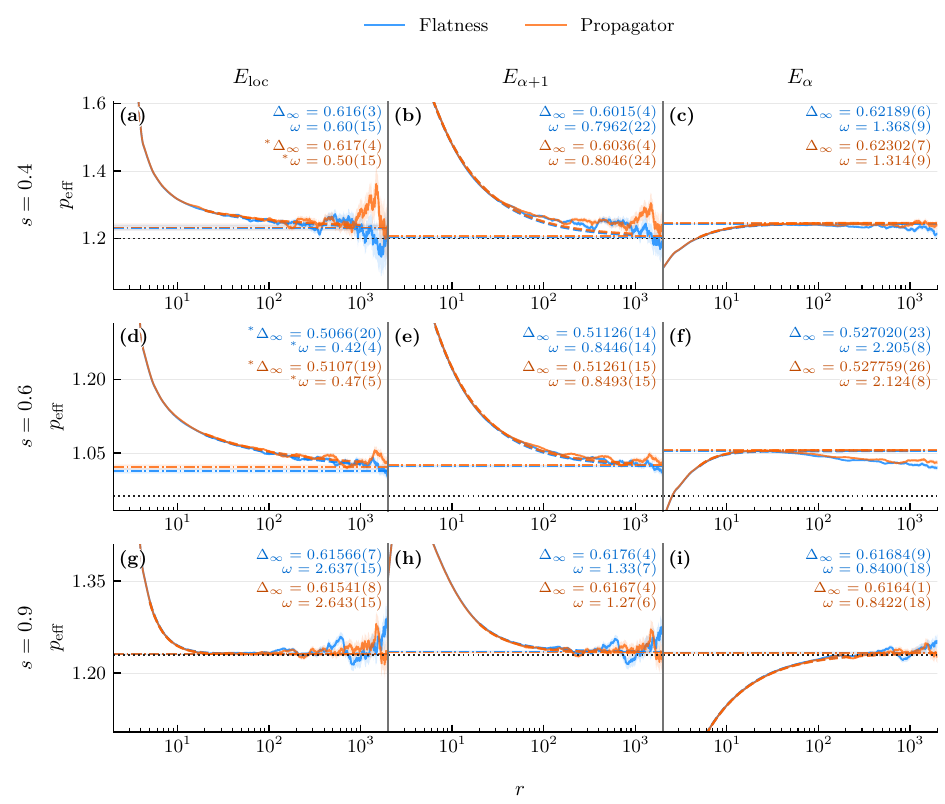}
    \caption{Comparison of the spin-flatness and propagator-crossing couplings at $L=10^6$, which we concluded was big enough in fig.~\ref{fig:energy-volume-comparison}.  For $s=0.4,0.6,0.9$, the respective couplings are $(0.181198,0.181158)$, $(0.281830,0.281804)$, and $(0.499740,0.499650)$. Rows, columns, fitting windows, and line conventions are the same as in fig.~\ref{fig:energy-volume-comparison}. Asterisks mark fits whose parameters are unstable under changes of the fitting window.}
    \label{fig:energy-coupling-comparison}
\end{figure}

We first assess finite-volume effects at fixed coupling. Figure~\ref{fig:energy-volume-comparison} shows no significant differences between $L=10^6$ and $L=10^7$, indicating that $L=10^6$ is sufficient at our current statistical precision. We then compare the couplings selected from the flatness of the spin correlator, the crossings of $F_L$, and the Binder threshold, discussed in Secs.~\ref{ssec:critfromprotect} and \ref{ssec:binder}. Figure~\ref{fig:energy-coupling-comparison} shows that, at $L=10^6$, the resulting differences in the extracted scaling dimensions are smaller than the statistical uncertainties.

The approach to asymptotic large distance behavior depends strongly on both the operator and $s$. The local energy generally approaches its extrapolated limit from above, consistent with contributions from higher-dimensional operators, while the matched-power energy approaches from below, consistent with the kernel-dependent corrections discussed previously. Among the operators considered, the steeper $E_{s+1}$ kernel generally provides the cleanest approach to asymptotic behavior, showing the importance of considering non-local operators and contrasting with the $\mathbb{Z}_2$ odd sector discussed above. 

The Gaussian point $s=0.4$, where $\Delta_{\phi^2}=1-s=0.6$, provides a useful control. At $L=10^6$, the finite-window extrapolations give $0.6015(4)$ for $E_{s+1}$, compared with $0.616(3)$ and $0.62189(6)$ for the local and matched-power energies, respectively. At $s=0.6$, the effective dimensions remain sensitive to the operator and fitting window. In contrast, at $s=0.9$ the different operators and coupling selections give consistent results, with approximately constant effective dimensions over one to two orders of magnitude in separation, compatible with the Pad\'e prediction. 

\subsection{Higher $\mathbb{Z}_2$-even operators}
\label{ssec:unprotected}

We next investigate whether higher dimension $\mathbb{Z}_2$-even scaling fields can be resolved using matrices of connected correlators \cite{jq2v-wjkf}. Although the nonlocal energy operators considered above can reduce corrections to the leading scaling dimension, their kernel-induced power laws complicate the identification of heavier operators.

Generally, a local even lattice operator overlaps with several continuum scaling fields,
\begin{equation}
    \mathcal O_a
    =c_{a,0}\mathbf 1+c_{a,2}\phi^2+c_{a,4}\phi^4
    +\text{descendants}+\cdots.
\end{equation}
Connected subtraction removes the identity contribution. At sufficiently large separations,
\begin{equation}
    \langle \mathcal{O}_{a,i}\mathcal{O}_{b,i+r}\rangle
    -\langle \mathcal{O}_a\rangle\langle \mathcal{O}_b\rangle\simeq
    c_{a,2}c_{b,2}\,A_2 r^{-2\Delta_{\phi^2}},
\end{equation}
which has rank one, for any operator overlapping with $\phi^2$. Thus, operators that are linearly independent on the lattice become effectively indistinguishable at long distances, making the extraction of subleading contributions particularly sensitive to signal to noise ratio.

We compared several operator bases and selected a local basis of twenty operators, including two-, four- and six-spin products with different spatial geometries. Nine of these consist of nonconstant, reflection-even spin combinations supported on five consecutive sites, constructed as
\begin{equation}
    \mathcal O_A(i)=\frac12\left[
    \prod_{a\in A}\sigma_{i+a}
    +\prod_{a\in A}\sigma_{i-a}
    \right],
    \qquad
    A\subset\{-2,-1,0,1,2\},\quad |A|=2,4,
\end{equation}
taking one representative from each reflection orbit $A\sim-A$. This gives six two-spin and three four-spin operators. We supplement them with all products of one, two or three centred bonds  $Q_a(i)=\sigma_{i-a}\sigma_{i+a}$, with $a=1,\ldots,4$. Removing the three combinations already present in the first set leaves eleven additional operators. All twenty operators are even under spin flip and reflection about their common centre, with support restricted to $\{i-4,\ldots,i+4\}$. Their support remains fixed as $L$ increases. A complementary approach is to optimize lattice representatives of continuum primary operators directly, as recently demonstrated using machine learning to reduce operator-dependent corrections to scaling in two-dimensional critical models~\cite{OppenheimGazitRingel2026}.

We construct the connected correlation matrix
\begin{equation}
    C_{ab}(r)=\frac1L\sum_i
    \langle\mathcal O_a(i)\mathcal O_b(i+r)\rangle
    -\langle\mathcal O_a\rangle
     \langle\mathcal O_b\rangle
\end{equation}
and solve the GEVP \cite{LUSCHER1990222,Blossier:2009kd}
\begin{equation}
    C(r)v_n(r)
    =\lambda_n(r,r_0)\,C(r_0)v_n(r),
    \qquad r_0=10.
\end{equation}
Since the operators become nearly linearly-dependent at large separations, the reference matrix can be ill-conditioned, amplifying statistical fluctuations in the subleading eigenvalues. We therefore regularize the problem in two steps. First, we diagonalize $C(0)$, retain eigenvalues bigger than $1\%$ of its largest eigenvalue, and rescale the retained eigenvectors so that the transformed $C(0)$ is the identity. We then diagonalize the transformed $C(r_0)$ and retain only its positive eigenvalues above $5\times 10^{-4} \lambda_{\rm max}$.

The retained subspace is fixed across separations, while the generalized eigenvectors are recomputed at each $r$. The principal correlators satisfy $\lambda_n(r_0,r_0)=1$ and, for isolated scaling fields, would behave as
\begin{equation}
    \lambda_n(r,r_0)
    =\left(\frac{r}{r_0}\right)^{-2\Delta_n}.
\end{equation}

\begin{figure}[htbp]
    \centering
    \includegraphics[width=\linewidth]
    {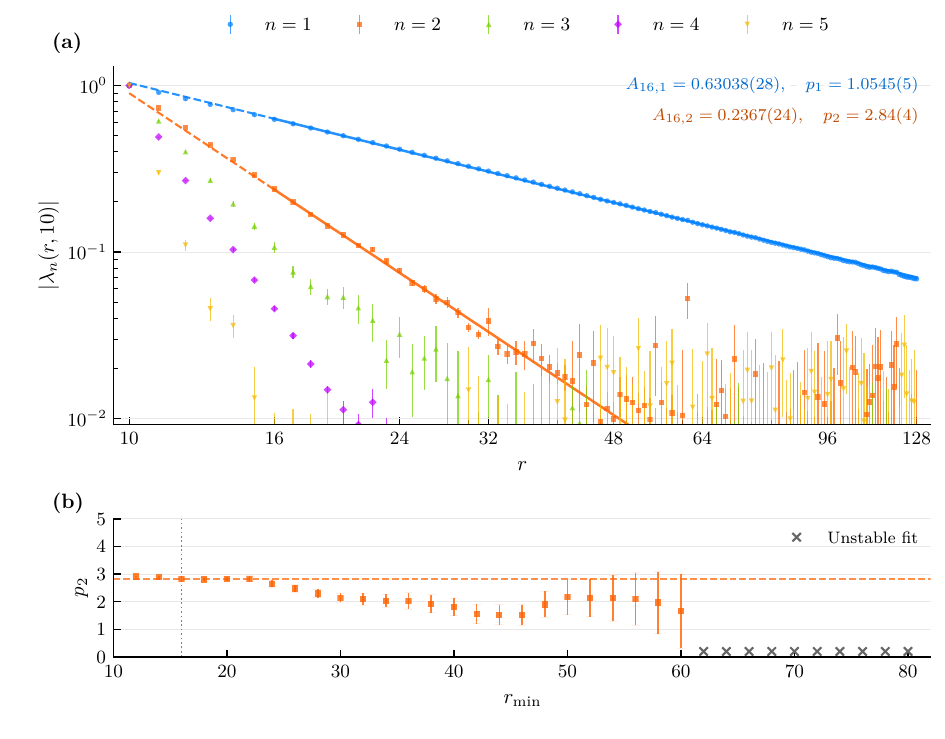}
    \caption{
    Generalized eigenvalue analysis of the twenty-operator local even basis described in the text, at $s=0.8$, $L=65536$ and $J_0=0.411149$, retaining five reference directions at $r_0=10$. Upper panel: correlator eigenvalues and power-law fits over $16\leq r\leq128$. Dashed segments extend the fits to smaller separations. Lower panel: second-mode power fitted over $r_{\min}\leq r\leq100$; the horizontal dashed line marks the $r_{\min}=16$ result. Grey crosses denote unstable fits whose values are not reported.
    }
    \label{fig:local-even-selected-gevp}
\end{figure}

Figure~\ref{fig:local-even-selected-gevp} shows a well-resolved leading mode and a more rapidly decaying sub-leading mode at $s=0.8$. Power-law fits yield $p_1=1.0545(5)$ over $16\leq r\leq128$ and $p_2=2.83(4)$ over $16\leq r\leq100$. However, the second exponent decreases to $p_2=2.0(2)$ when the lower cut is raised to $r_{\min}=32$, and becomes poorly constrained at larger cuts. Thus, although a subleading signal is resolved, there is still some dependence on the fit window. 

\subsection{Three-point functions}
\label{ssec:three-point}

Having determined the leading even scaling dimension, we turn to the mixed three-point function $\langle\phi\phi\phi^2\rangle$ to extract the OPE coefficient $C_{\phi\phi\phi^2}$. For unit-normalized continuum operators, conformal invariance fixes
\begin{equation}
    \langle\phi(x_1)\phi(x_2)\phi^2(x_3)\rangle
    =
    \frac{C_{\phi\phi\phi^2}}
    {|x_{12}|^{2\Delta_\phi-\Delta_{\phi^2}}
     |x_{13}|^{\Delta_{\phi^2}}
     |x_{23}|^{\Delta_{\phi^2}}}.
    \label{eq:local-three-point-ope}
\end{equation}
In the Gaussian regime, Wick contractions give $C_{\phi\phi\phi^2}=\sqrt{2}$, where the unit-normalized even operator is $:\!\phi^2\!:/\sqrt{2}$. In the interacting regime, the coefficient is dynamical, with perturbative predictions available to two orders near $s=1$~\cite{BenedettiLauriaMazacVanVliet2026} and one order near $s=1/2$ \cite{BehanLauriaNocchiVanVliet2024}.

Monte Carlo measurements of critical two- and three-point functions have previously been used to extract OPE coefficients in the three-dimensional Ising universality class~\cite{PhysRevE.97.012119}. On the lattice, we use the local energy operator $E_i=E_{\mathrm{loc},i}$ defined in eq.~\eqref{eq:energy-operators}. For symmetric insertions at $-x$, $0$ and $x$, we define
\begin{align}
    G_\sigma(r)
    &=\langle\sigma_0\sigma_r\rangle,\\
    C_E(r)
    &=\langle E_0E_r\rangle-\langle E\rangle^2,\\
    T(x)
    &=\langle\sigma_{-x}\sigma_x E_0\rangle
      -G_\sigma(2x)\langle E\rangle.
\end{align}
The subtraction removes the identity contribution from the even insertion. Assuming the leading overlap of $E$ is with $\phi^2$, the ratio
\begin{equation}
    C_{\phi\phi\phi^2}^{\mathrm{eff}}(x)
    =
    \frac{T(x)\sqrt{C_E(2x)}}
    {G_\sigma(2x)C_E(x)}
    \longrightarrow C_{\phi\phi\phi^2},
    \qquad 1\ll x\ll L,
    \label{eq:symmetric-ope-ratio}
\end{equation}
cancels both lattice normalization amplitudes and the leading scaling powers. Thus, the coefficient can be extracted without an independent determination of $\Delta_{\phi^2}$. We choose the sign of $\phi^2$ so that its overlap with the local energy is positive.

We also consider asymmetric configurations, placing the spins at $-r$ and $q$, with the energy insertion at the origin. Defining
\begin{equation}
    T(r,q)
    =\langle\sigma_{-r}\sigma_q E_0\rangle
     -G_\sigma(r+q)\langle E\rangle,
\end{equation}
we obtain
\begin{equation}
    C_{\phi\phi\phi^2}^{\mathrm{eff}}(r,q)
    =
    \frac{T(r,q)}{G_\sigma(r+q)}
    \sqrt{\frac{C_E(r+q)}{C_E(r)C_E(q)}}.
    \label{eq:asymmetric-ope-ratio}
\end{equation}
Conformal invariance predicts that this ratio approaches the same constant independently of the insertion geometry. Comparing symmetric and asymmetric configurations therefore provides an additional test of the conformal three-point structure and its finite-distance corrections.

We study $s=0.9$ and the Gaussian control $s=0.4$ at $L=10^6$. Guided by the scaling window of the local energy correlator, we restrict all pairwise separations to exceed $10^2$, with symmetric configurations spanning $128\leq x\leq4096$. We compare different insertion ratios $r:q$ and vary the lower distance cut to assess corrections to scaling.

\begin{figure}[t]
    \centering
    \includegraphics[width=\textwidth]{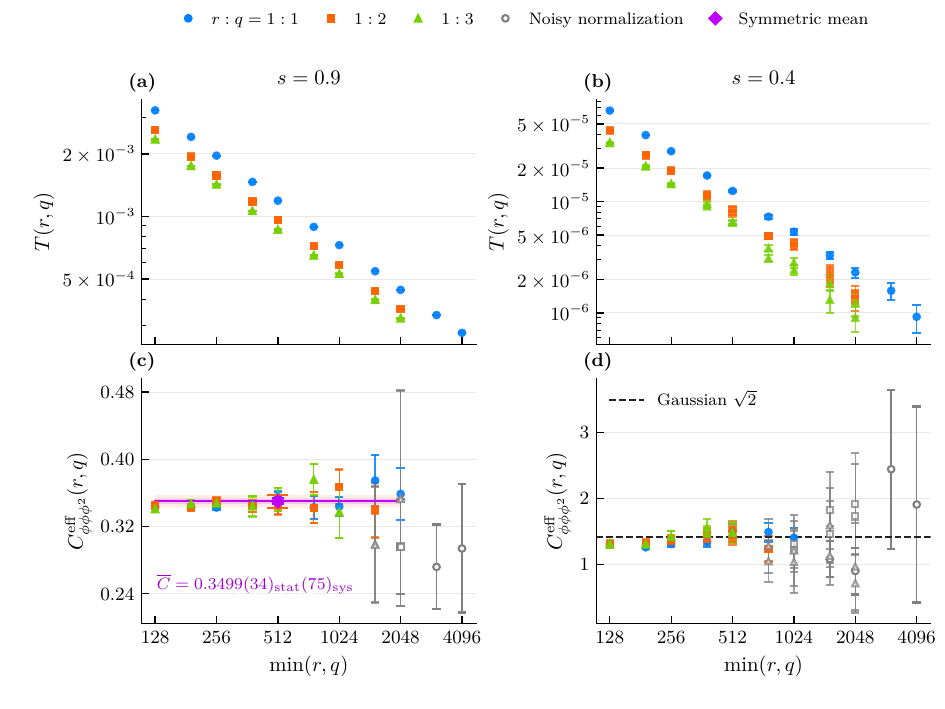}
    \caption{
        Connected three-point function $T(r,q)$ (top) and effective OPE coefficient $C_{\phi\phi\phi^2}^{\mathrm{eff}}(x)$ (bottom) at $L=10^6$, for $s=0.9$ (left) and $s=0.4$ (right). Each point corresponds to one measured separation. In the $s=0.9$ panel, the purple diamond denotes the unweighted average over $128\leq x\leq2048$. The ratios $1:2$ and $1:3$ refer to the distances of the two spins from the energy insertion: with the energy at the origin, they correspond to spins at $(-r,2r)$ and $(-r,3r)$, respectively. Both reflected orientations, obtained by interchanging $r$ and $q$, are included. The asymmetric geometries provide a consistency check and are not included in the reported average or either uncertainty estimate. For the extracted OPE coefficient, the inner and outer error bars show the statistical and systematic uncertainties, respectively. Hollow points indicate poorly resolved two-point normalization. The dashed line is the Gaussian value $\sqrt{2}$.
    }
    \label{fig:three-point-ope}
\end{figure}

The connected three-point function, panel $a)$ and $b)$ of fig. \ref{fig:three-point-ope} is well resolved over the measured distance range for both values of $s$. At $s=0.9$, the effective OPE coefficient is approximately independent of separation within the statistical uncertainties. Taking the unweighted average of the nine symmetric estimates with $128\leq x\leq2048$, we obtain
\[
    \overline{C}_{\phi\phi\phi^2}^{\mathrm{eff}}
    =0.3499\pm0.0034_{\mathrm{stat}}\pm0.0075_{\mathrm{sys}}.
\]
We estimate the systematic uncertainty as $\max_{x<512}|C_{\phi\phi\phi^2}^{\mathrm{eff}}(x) -\overline{C}_{\phi\phi\phi^2}^{\mathrm{eff}}|$ over the measured separations. This can be compared with a naive ressumation of the estimate from the expansion around the short-range end
\begin{equation}
    |C_{\phi\phi\phi^2}^{\rm{SR}}|=\frac{\sqrt{1-\epsilon}}{2}+\frac{15}{32}(1-\epsilon) +\left(\frac{543}{1024}-\frac{\pi^2}{24}\right)(1-\epsilon)^{3/2}+O((1-\epsilon)^2)\approx 0.328 
\end{equation}
and from the mean-field end
\begin{equation}
    C_{\phi\phi\phi^2}^{\rm{MF}}= \sqrt{2}-1.39399 \epsilon + O(\epsilon^2) \approx0.299
\end{equation}
Note that the sign of a single OPE coefficient cannot be unambiguously fixed due to a possible overall sign redefinition of all local operators. In fact, the short-range result presented in REF has a negative sign. 
\medskip

In the Gaussian regime, the decay of $T(x)$ is consistent with the expected $x^{-1.2}$ behavior, while the normalized coefficient shows a short-distance offset from $\sqrt{2}$ and increasingly large uncertainties at long distances. At large separations, the connected energy two-point functions entering
the normalization become poorly resolved, limiting the precision of
the effective OPE coefficient even where the connected three-point
function remains statistically resolved.

\subsection{Four-point functions}
\label{ssec:four-point}

The four-point function of the spin field provides a complementary route
to the even-sector OPE data, without introducing an explicit lattice
energy operator. For ordered insertions $0<x<y<D$, we define
\begin{equation}
    R_L(x,y,D)
    =
    \frac{\langle\sigma_0\sigma_x\sigma_y\sigma_D\rangle}
         {G_\sigma(x)G_\sigma(D-y)},
    \qquad
    z=\frac{x(D-y)}{y(D-x)}\,,
    \label{eq:four-point-ratio}
\end{equation}
where $z$ is the standard conformally invariant cross-ratio. The spin normalization cancels in this ratio $R_L$. When all insertion
separations are large compared with the lattice spacing and $D\ll L$,
conformal invariance predicts that $R_L$ approaches a function $R(z)$ of
the cross ratio alone. Its conformal-block expansion is
\begin{equation}
    R(z)-1
    =
    \sum_{\substack{\mathcal O\in\phi\times\phi\\
                    \mathcal O\ne\mathbf 1}}
    C_{\phi\phi\mathcal O}^{\,2}\,
    g_{\Delta_{\mathcal O}}(z),
    \qquad
    g_\Delta(z)
    =
    z^\Delta\,{}_2F_1(\Delta,\Delta;2\Delta;z).
    \label{eq:four-point-block-expansion}
\end{equation}
The subtraction removes the identity contribution in this channel.
Each block includes the full descendant series of an exchanged primary.
The leading nonidentity contribution comes from $\phi^2$, allowing its
dimension and OPE coefficient to be extracted from spin measurements
alone.

We consider $s=0.9$ at $L=10^5$, fixing the outer separation to $D=512$
and varying the positions of the two interior spins. We compare
\begin{equation}
    (0,a,D-a,D),
    \qquad
    (0,a,D-2a,D),
    \label{eq:four-point-geometries}
\end{equation}
whose outer-pair separations have ratios $1:1$ and $1:2$, respectively.
Reflected configurations are averaged together. These two geometries
probe the same conformal correlation function while retaining a priori different
finite-distance corrections.

\begin{figure}[t]
    \centering
    \includegraphics[width=\textwidth]
        {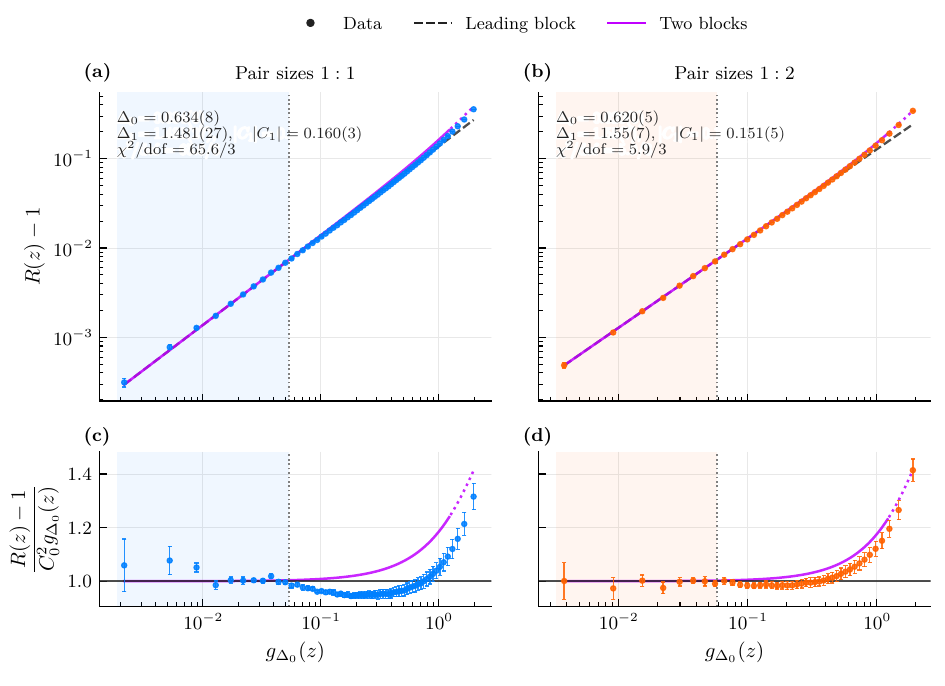}
    \caption{
        Spin four-point function at $s=0.9$, $L=10^5$ and $D=512$. Left and right columns correspond to outer-pair separation ratios $1:1$ and $1:2$. Upper panels show the measured excess $R_L-1$, against the full leading conformal block. Dashed curves are the leading-block fits obtained in the shaded region $z\leq10^{-2}$. Purple curves add a second full block, fitted at $10^{-2}<z\leq0.75$ while retaining the leading parameters from the small-$z$ fit; the dotted line shows the extrapolation to $z>0.75$. Lower panels show $[R(z)-1]/[C_0^2g_{\Delta_0}(z)]$, so that deviations from unity measure departures from the leading-block prediction. Purple curves include the second block. Points show the unbinned data; fits use binned correlators and their covariance. Errors retain the paired uncertainty of the data and fitted normalization.
    }
    \label{fig:four-point-block-coordinate}
\end{figure}

At small cross ratio, the leading conformal block dominates,
\begin{equation}
    R_L-1
    \simeq C_0^2g_{\Delta_0}(z),
    \qquad
    \Delta_0\longrightarrow\Delta_{\phi^2},
    \qquad
    |C_0|\longrightarrow|C_{\phi\phi\phi^2}|.
    \label{eq:four-point-leading-fit}
\end{equation}
We fit the dimension and amplitude independently for each geometry,
using $z\leq10^{-2}$. Figure~\ref{fig:four-point-block-coordinate} shows
the measured excess against the resulting full-block coordinate.
Unlike a plot against $z^{\Delta_0}$, this representation includes the
descendant contribution, so a single conformal family gives a straight
line even beyond the asymptotically small-$z$ regime. The lower panels
divide out the fitted leading contribution, making departures from this
description more visible.

The leading estimates are given in Table \ref{tab:four-point-leading}.
\begin{table}[t]
    \centering
    \begin{tabular}{cccc}
        \toprule
        Outer-pair ratio
        & $\Delta_0$
        & $|C_0|$
        & $\chi^2/\mathrm{dof}$ \\
        \midrule
        $1:1$ & $0.634(8)$ & $0.371(5)$ & $7.4/3$ \\
        $1:2$ & $0.620(5)$ & $0.357(4)$ & $5.7/3$ \\
        \bottomrule
    \end{tabular}
    \caption{
        Leading even-channel parameters extracted from the spin
        four-point function at $s=0.9$, $L=10^5$ and $D=512$.
        Parentheses denote paired jackknife statistical uncertainties.
        These are finite-distance estimates; the errors exclude
        coupling, volume and fit-window dependence.
    }
    \label{tab:four-point-leading}
\end{table}
The unequal-pair estimate for the OPE coefficient is close to the three-point determination in
Sec.~\ref{ssec:three-point}, while the equal-pair geometry gives a higher
central value. Their difference illustrates the residual dependence on
the insertion geometry. 

At larger cross ratio, the measured excess rises above the leading-block
prediction. To estimate the next contribution, we retain the leading fit
and describe the remaining signal by a second full block,
\begin{equation}
    R_L-1
    \simeq
    C_0^2g_{\Delta_0}(z)
    +C_1^2g_{\Delta_1}(z),
    \qquad \Delta_1>\Delta_0.
    \label{eq:four-point-two-block-fit}
\end{equation}
The second amplitude is fitted without a positivity constraint. For the $1:2$ geometry, the fitted amplitude is positive and gives
\begin{equation}
    \Delta_1^{\mathrm{eff}}=1.55(7),
    \qquad
    |C_1^{\mathrm{eff}}|=0.151(5),
    \qquad
    \chi^2/\mathrm{dof}=5.9/3.
    \label{eq:four-point-subleading-estimate}
\end{equation}
This provides a tentative estimate of the subleading exchange. Both results compare well to a naive ressumation of the short-range expansion. Indeed \eqref{phi4SR} gives
\begin{equation}
    \Delta_1^{\rm{SR}}\approx 1.472\,,
\end{equation}
while the prediction for the  subleading OPE coefficient is
\begin{equation}
        |C_{\phi\phi\phi^4}^{\rm{SR}}|=\frac{\sqrt{1-\epsilon}}{2}-\frac{15}{32}(1-\epsilon) +\left(\frac{543}{1024}-\frac{\pi^2}{24}\right)(1-\epsilon)^{3/2}+O((1-\epsilon)^2)\approx 0.141\,.
\end{equation}

The equal-pair geometry is less well described: its second-block fit
has $\chi^2/\mathrm{dof}=65.6/3$, and we do not interpret its fitted
parameters as a determination of another operator.

The lower panels of fig. \ref{fig:four-point-block-coordinate} clarify the limitation of these estimates. A positive second block can
describe the upward departure at larger cross ratio, but cannot account
for the intermediate deficit below the leading prediction.
Finite-distance corrections to the leading contribution, as well as
further exchanged operators, can affect this region.
Thus Eq.~\eqref{eq:four-point-subleading-estimate} characterizes an
effective subleading contribution at the measured scale, rather than cleanly
isolating a unique continuum primary. The leading dimension and OPE
coefficient are more directly accessible, while a controlled extraction
of the next operator requires greater stability under changes of the
insertion scale and fit window.

\section{Summary of results and comparison with Pad\'e}
\label{sec:comparison}

\begin{figure}[t]
    \centering
    \includegraphics[width=\linewidth]
    {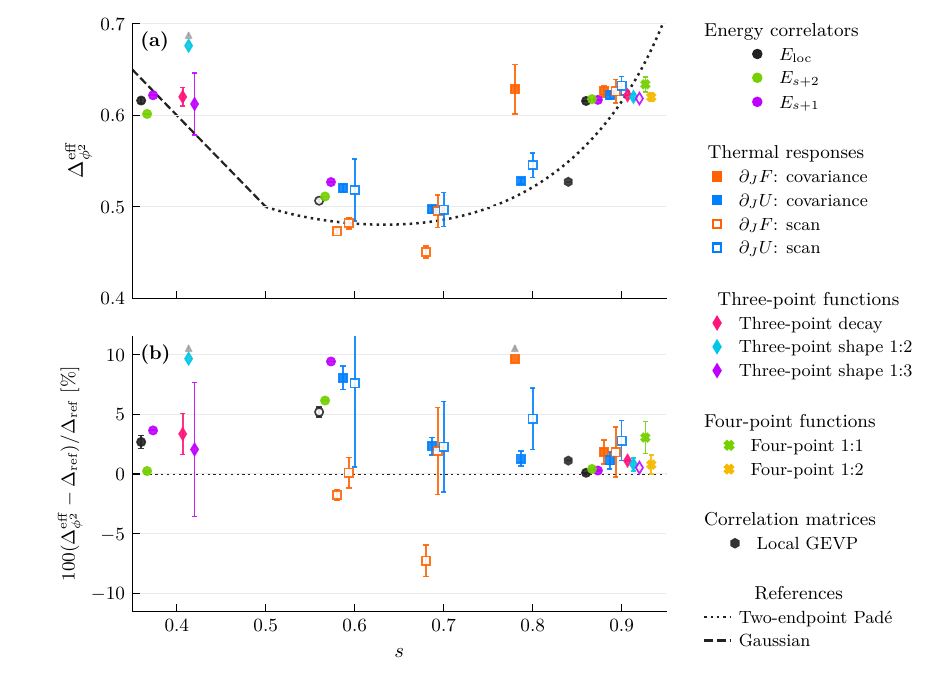}
    \caption{Estimates of the leading $\mathbb{Z}_2$ even dimension from the different analyses. (a) Extracted dimensions, compared with the Gaussian result and the two-endpoint Pad\'e interpolation. (b) Percentage deviations from the corresponding reference. \textbf{Horizontal offsets separate symbols for clarity.} Gray arrows indicate estimates outside the displayed range; their vertical positions are not quantitative.}
    \label{fig:conformal-dimensions-comparison}
\end{figure}

Figure~\ref{fig:conformal-dimensions-comparison} collects our estimates of $\Delta_{\phi^2}$ from thermal responses, connected two-point functions, three-point functions, four-point functions, and correlation matrices. We compare them with the two-endpoint Pad\'e interpolation constructed from the perturbative expansions near $s=1/2$ and $s=1$. This interpolation provides a useful reference, but should not be interpreted as an exact result with a controlled uncertainty. In the Gaussian regime, the reference is the exact dimension $\Delta_{\phi^2}=1-s$.

The different methods reproduce the overall dependence on $s$, although significant differences remain between observables and analysis choices. At $s=0.4$, the steeper nonlocal energy correlator gives a value close to the exact Gaussian dimension, while other estimators exhibit visible finite-distance or finite-size deviations. These effects are particularly apparent at intermediate $s$, where the thermal responses and the different energy operators do not always approach a common estimate within their quoted statistical errors. The agreement is clearest at $s=0.9$: thermal responses, energy correlators, and three- and four-point analyses cluster within a few percent of the Pad\'e reference. Their spread nevertheless shows that the uncertainties associated with the critical coupling, finite-size corrections, operator choice, and fitting window cannot be inferred from statistical errors alone. 

An interesting feature of our results is that $s=0.9$ is systematically easier to analyze than the other values of $s$ considered. The correlators exhibit longer and more stable scaling plateaus, different operators give more consistent estimates, and the extracted conformal data show smaller apparent systematic deviations. Having a small value of $\Delta_\phi$ while simultaneously having relatively large values of $\Delta_{\phi^2},\Delta_{\phi^3}$ and $\Delta_{\phi^4}$ presumably helps stabilize the results. However, understanding this difference in more detail could help identify more effective observables and analysis strategies throughout the family of LRI CFTs.

\begin{figure}[t]
    \centering
    \includegraphics[width=\linewidth]
    {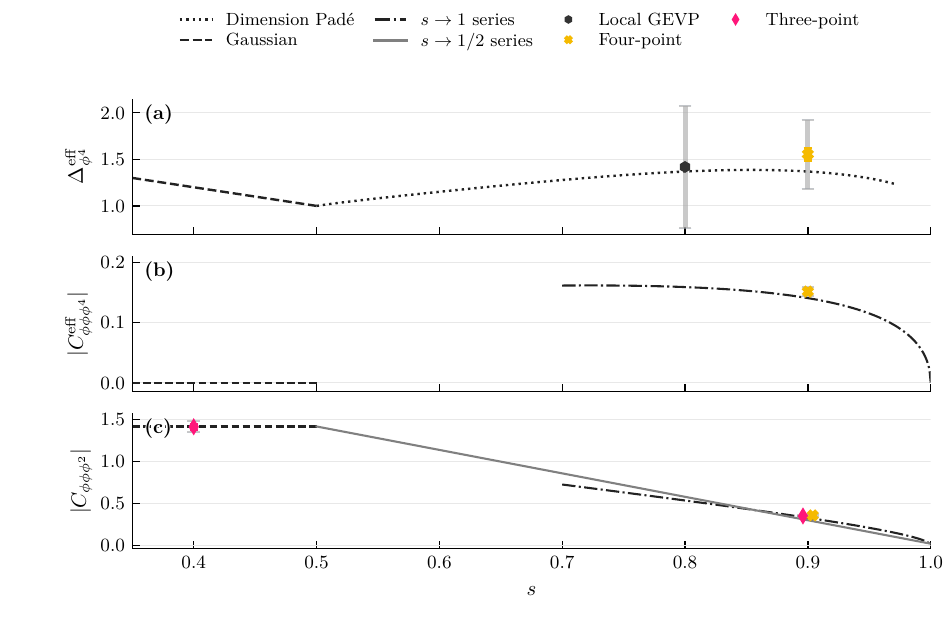}
    \caption{Scaling dimensions and OPE coefficients from higher-point and correlation-matrix analyses. (a) Effective candidates for the subleading even dimension, from the second GEVP mode at $s=0.8$ and the next four-point block at $s=0.9$, compared with the $[3,3]$ two-endpoint interpolation for $\Delta_{\phi^4}$. (b) Magnitude of the corresponding effective four-point OPE coefficient and comparison to the short-range expansion. (c) Leading OPE coefficient $|C_{\phi\phi\phi^2}|$ from three- and four-point functions, compared with the Gaussian value and perturbative endpoint expansions. Colored error bars represent statistical uncertainties; gray bars indicate variations under the analysis choices considered. The identification of the subleading contributions with $\phi^4$ remains tentative.}
    \label{fig:conformal-subleading-dimensions}
\end{figure}

Figure~\ref{fig:conformal-subleading-dimensions} extends the comparison to OPE data and the subleading even sector. The Gaussian three-point result is close to $|C_{\phi\phi\phi^2}|=\sqrt{2}$, while at $s=0.9$ the three-point and four-point measurements give comparable leading OPE coefficients. The differences between the two four-point geometries and the three-point estimate are larger than some of their statistical errors, again indicating residual finite-distance systematics. The normalized three-point function provides an estimate of the OPE coefficient independent of $\Delta_{\phi^2}$, while the spin four-point function allows both quantities to be extracted independently, with different sources of systematic uncertainty.

The subleading sector is less well controlled. At $s=0.8$, the second GEVP mode has an effective dimension of the order expected for $\phi^4$, but its fitted exponent changes substantially with the distance window. At $s=0.9$, a two-block fit to one four-point geometry gives an effective subleading dimension and coefficient near the perturbative value. These measurements therefore indicate sensitivity to subleading contributions but do yet not establish an unambiguous determination of $\Delta_{\phi^4}$ or its OPE coefficient. Given the limited CPU time devoted to these exploratory measurements and the efficiency of the algorithm, substantially increasing the statistics is readily feasible. We expect that this will allow a more controlled extraction of the subleading conformal data, including a systematic assessment of finite-distance corrections and conformal-block truncation effects.

\begingroup\small
\setlength{\tabcolsep}{3pt}
\setlength{\LTcapwidth}{\textwidth}
\begin{longtable}{clrrrrr}
\caption{Summary of the conformal dimensions and OPE coefficients. Gray rows give Gaussian/two-endpoint Pad\'e or truncated endpoint-series references.  \textit{Estimated} CPU are orders of magnitude in core-hours.}
\label{tab:conformal-data-summary}
\label{tab:conformal-dimensions-summary}
\label{tab:conformal-subleading-summary}\\
\toprule
$s$ & Analysis & $\Delta_{\phi^2}^{\rm eff}$ & $\Delta_{\phi^4}^{\rm eff}$ & $|C_{\phi\phi\phi^2}|$ & $|C_{\phi\phi\phi^4}^{\rm eff}|$ & CPU (core-h) \\
\midrule
\endfirsthead
\toprule
$s$ & Analysis & $\Delta_{\phi^2}^{\rm eff}$ & $\Delta_{\phi^4}^{\rm eff}$ & $|C_{\phi\phi\phi^2}|$ & $|C_{\phi\phi\phi^4}^{\rm eff}|$ & CPU (core-h) \\
\midrule
\endhead
\midrule
\multicolumn{7}{r}{Continued on next page}\\
\endfoot
\bottomrule
\endlastfoot
\rowcolor{gray!12} 0.4 & Gaussian reference & $0.6$ & -- & $\simeq 1.41421$ & -- & -- \\
0.4 & Energy, $E_{\rm loc}$ & $0.616(13)$ & -- & -- & -- & $10^{4}$ \\
0.4 & Energy, $E_{s+2}$ & $0.6015(62)$ & -- & -- & -- & $10^{4}$ \\
0.4 & Energy, $E_{s+1}$ & $0.6219(36)$ & -- & -- & -- & $10^{4}$ \\
0.4 & Three-point decay & $0.620(14)$ & -- & -- & -- & $10^{3}$ \\
0.4 & Three-point shape $1{:}2$ & $0.79(25)$ & -- & -- & -- & $10^{3}$ \\
0.4 & Three-point shape $1{:}3$ & $0.61(10)$ & -- & -- & -- & $10^{3}$ \\
0.4 & Three-point ratio & -- & -- & $1.410(75)$ & -- & $10^{3}$ \\
\midrule
\rowcolor{gray!12} 0.6 & Pad\'e reference & $\simeq 0.48163$ & -- & -- & -- & -- \\
0.6 & Energy, $E_{\rm loc}$ & $0.5066(45)$ & -- & -- & -- & $10^{4}$ \\
0.6 & Energy, $E_{s+2}$ & $0.5113(82)$ & -- & -- & -- & $10^{4}$ \\
0.6 & Energy, $E_{s+1}$ & $0.5270(38)$ & -- & -- & -- & $10^{4}$ \\
0.6 & Covariance, $F$ & $0.473(14)$ & -- & -- & -- & $10^{2}$ \\
0.6 & Covariance, $U$ & $0.5204(79)$ & -- & -- & -- & $10^{2}$ \\
0.6 & Coupling scan, $F$ & $0.4822(61)$ & -- & -- & -- & $10^{3}$ \\
0.6 & Coupling scan, $U$ & $0.518(34)$ & -- & -- & -- & $10^{3}$ \\
\midrule
\rowcolor{gray!12} 0.7 & Pad\'e reference & $\simeq 0.48589$ & -- & -- & -- & -- \\
0.7 & Covariance, $F$ & $0.451(26)$ & -- & -- & -- & $10^{2}$ \\
0.7 & Covariance, $U$ & $0.4972(37)$ & -- & -- & -- & $10^{2}$ \\
0.7 & Coupling scan, $F$ & $0.495(18)$ & -- & -- & -- & $10^{3}$ \\
0.7 & Coupling scan, $U$ & $0.497(18)$ & -- & -- & -- & $10^{3}$ \\
\midrule
\rowcolor{gray!12} 0.8 & Pad\'e reference & $\simeq 0.52143$ & $\simeq 1.36880$ & -- & -- & -- \\
0.8 & Covariance, $F$ & $0.628(53)$ & -- & -- & -- & $10^{2}$ \\
0.8 & Covariance, $U$ & $0.5281(40)$ & -- & -- & -- & $10^{2}$ \\
0.8 & Coupling scan, $U$ & $0.546(13)$ & -- & -- & -- & $10^{3}$ \\
0.8 & Local GEVP & $0.527(10)$ & $1.42(66)$ & -- & -- & $10^{3}$ \\
\midrule
\rowcolor{gray!12} 0.9 & Pad\'e reference & $\simeq 0.61506$ & $\simeq 1.36872$ & -- & -- & -- \\
\rowcolor{gray!12} 0.9 & $s\to1$ series & -- & -- & $\simeq 0.32800$ & $\simeq 0.14050$ & -- \\
\rowcolor{gray!12} 0.9 & $s\to1/2$ series & -- & -- & $\simeq 0.29902$ & -- & -- \\
0.9 & Energy, $E_{\rm loc}$ & $0.6157(11)$ & -- & -- & -- & $10^{4}$ \\
0.9 & Energy, $E_{s+2}$ & $0.6176(29)$ & -- & -- & -- & $10^{4}$ \\
0.9 & Energy, $E_{s+1}$ & $0.6168(21)$ & -- & -- & -- & $10^{4}$ \\
0.9 & Covariance, $F$ & $0.6264(93)$ & -- & -- & -- & $10^{2}$ \\
0.9 & Covariance, $U$ & $0.6221(45)$ & -- & -- & -- & $10^{2}$ \\
0.9 & Coupling scan, $F$ & $0.626(13)$ & -- & -- & -- & $10^{3}$ \\
0.9 & Coupling scan, $U$ & $0.632(10)$ & -- & -- & -- & $10^{3}$ \\
0.9 & Three-point decay & $0.6219(98)$ & -- & -- & -- & $10^{3}$ \\
0.9 & Three-point shape $1{:}2$ & $0.6200(64)$ & -- & -- & -- & $10^{3}$ \\
0.9 & Three-point shape $1{:}3$ & $0.618(20)$ & -- & -- & -- & $10^{3}$ \\
0.9 & Four-point $1{:}1$ & $0.634(24)$ & -- & -- & -- & $10^{1}$ \\
0.9 & Four-point $1{:}2$ & $0.620(15)$ & $1.55(38)$ & $0.357(14)$ & $0.1513(93)$ & $10^{1}$ \\
0.9 & Three-point ratio & -- & -- & $0.3499(82)$ & -- & $10^{3}$ \\
\end{longtable}
\endgroup

\section{Discussion}
\label{sec:discuss}
The versatility of the one-dimensional long range Ising model as a testbed for non-perturbative methods in CFT has hopefully been made clear by our explorations. We computed the scaling dimension of the energy operator in multiple different ways, from thermodynamic response, two-point functions, three-point and even-four point functions; we obtained estimates for OPE coefficients and for the full cross-ratio dependence of the four-point function of the order parameter. All these measurements are made possible by the small conformal dimension of the magnetic field operator and from the powerful $O(L)$ FT algorithm. 

Our results are not meant to be taken as the definitive estimates for any of these quantities, rather they aim to be the starting point for subsequent high-precision studies, with more controlled finite volume corrections targeted at each of the observables. In fact, most of the observables we studied (excluding the two-point functions of the energy operator) required only very modest CPU time usage and more refined studies would therefore be rather straightforward to pursue.

There are many interesting open targets to further expand the breadth of the 1d LRI as a playground for CFT. First, a controlled construction of lattice operators whose dominant continuum contribution are subleading operators, both in the $\mathbb{Z}_2$ even and $\mathbb{Z}_2$ odd sectors opens the door for the measurement of many interesting observables. For example, in the $\mathbb{Z}_2$ odd sector, the two leading operators $\phi$ and $\phi^3$ which form a shadow pair, are meant to satisfy highly non-trivial OPE relations of the form \cite{PaulosRychkovVanReesZan2016,Behan:2018hfx,BehanDiPietroLauriaVanRees2020,LauriaLiendoVanReesZhao2021,BehanLauriaNocchiVanVliet2024}
\begin{equation}
   \frac{C_{\phi ij}C_{\phi^3kl}}{C_{\phi kl}C_{\phi^3 ij}} = \rm fixed
\end{equation}
where $i,j,k,l$ are arbitrary local operators. Verifying these predictions non-perturbatively is a very interesting benchmark for testing control over the subleading operator $\phi^3$.

In the $\mathbb{Z}_2$ even sector, the non-local construction of the energy operator was key in producing some of the most reliable estimates, but introduces contamination via additional power laws which appear from overlapping terms in the sums. Estimating the energy-energy-energy OPE coefficient $C_{\phi^2\phi^2\phi^2}$ will therefore require a more fine-grained understanding of the construction, including how to systematically subtract the spurious power laws. Furthermore, determining the scaling dimension and OPE coefficients of the subleading operator $\phi^4$ will require not only control over the non-local smearing as well as the ability to cleanly construct orthogonal linear combinations of such objects.

Finally, in the CFT literature, a strong emphasis is placed on the determination of many operator dimensions, including several excited states, far beyond the operators that control traditional critical exponents. Our GEVP analysis included twenty operators and seemed to isolate, at least qualitatively, 5 different power laws in the $\mathbb{Z}_2$ even sector. Is it possible to push these studies to high-precision and turn Monte Carlo into a realistic conformal spectrometer for the 1d LRI model?

\paragraph{Note added}
We coordinated the submission of this article with that of \cite{ZhengLongRangeIsingWIP} by Zechuan Zheng, which performed a vast study of the LRI model in $d=1,2,3$ using various perturbative and non-perturbative methods, including Monte Carlo simulations.

\paragraph{AI disclosure}
We used ChatGPT Sol 6.1 and Astra 6 (OpenAI) to assist in code writing, data analysis,  plot generation and some mild text editing. The authors take full responsibility for the validity of the results and for their scientific interpretation.

\paragraph{Acknowledgements}
We thank Miguel Paulos for useful discussions. AA further thanks Andreia Gon\c{c}alves for continued support and inspiration. AA is funded by the European Union (ERC, FUNBOOTS, project number 101043588). Views and opinions  expressed are however those of the author(s) only and do not necessarily reflect those of  the European Union or the European Research Council Executive Agency. Neither the European Union nor the granting authority can be held responsible for them. JM was supported by FCT - Funda\c{c}\~ao para a Ci\^encia e Tecnologia, I.P., through doctoral grant 2021.04743.BD \href{https://doi.org/10.54499/2021.04743.BD}{https://doi.org/10.54499/2021.04743.BD}. JM acknowledges computing time provided by GRID FEUP. JM also thanks the Centro de F\'isica do Porto, funded by the Portuguese Foundation for Science and Technology (FCT) through Strategic Funding UIDB/04650/2020.

\bibliography{Bib.bib}
\bibliographystyle{JHEP}

\flushbottom

\appendix

\section{Critical couplings at all $s$}\label{app:all critical couplings finite volume dependence}
\begin{figure}[h]
    \centering
    \includegraphics[width=0.82\linewidth]{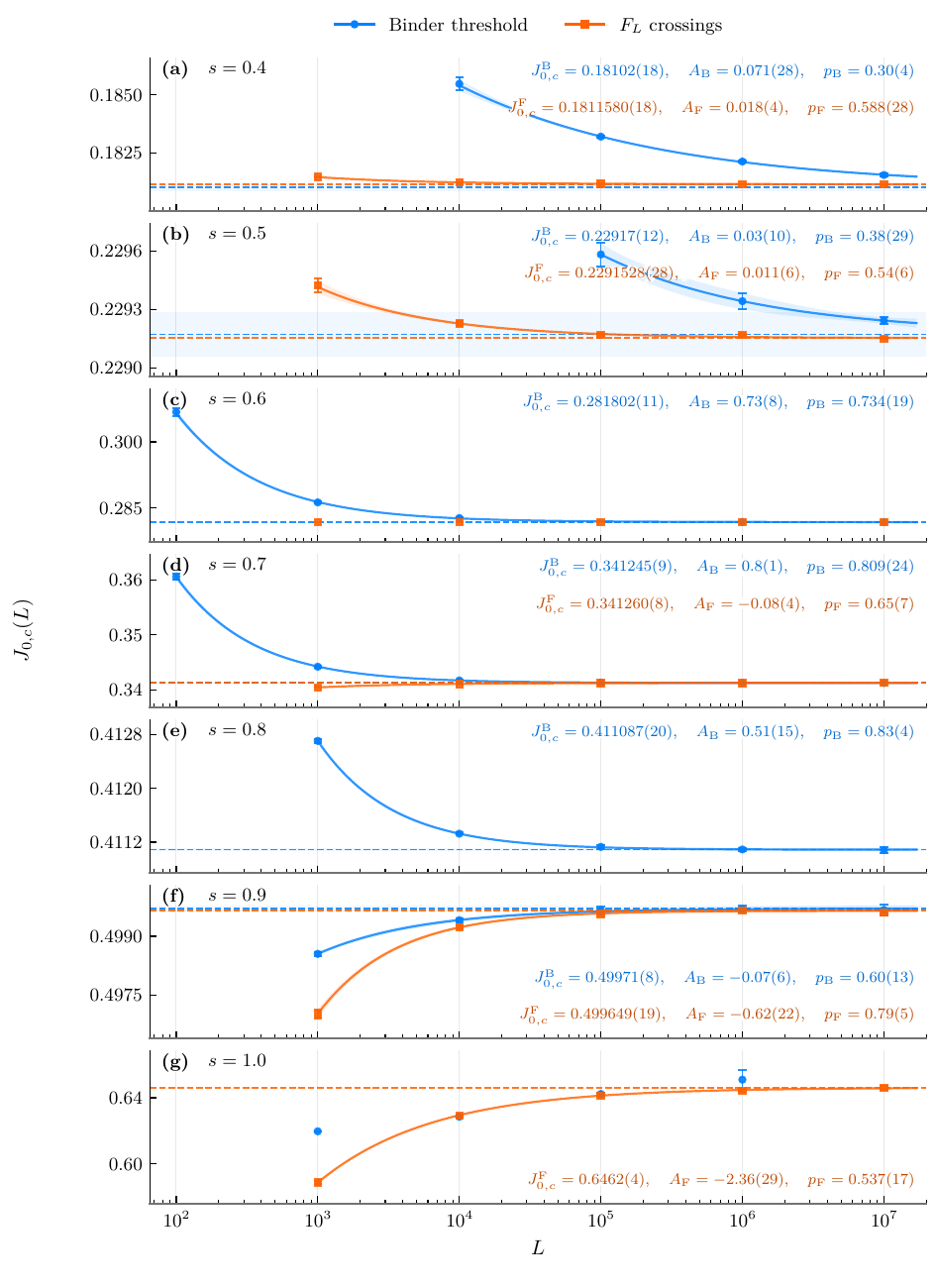}
    \caption{Finite-size critical-coupling estimates at several values of $s$ from the Binder threshold (blue circles) and consecutive-size crossings of $F_L$ (orange squares). The x-axis coordinate is the larger size in each pair. Solid curves fit $J_{0,c}(L)=J_{0,c}+A L^{-p}$ and dashed lines represent the extrapolated values.  One-standard-error uncertainties are represented by bands.}
    \label{fig:critical-couplings-all-s}
\end{figure}

Figure~\ref{fig:critical-couplings-all-s} collects the finite-size critical-coupling estimates for $s=0.4,0.5,\ldots,1.0$, comparing the Binder threshold $U_L=0.1$ with consecutive-size crossings of $F_L$. 

\end{document}